%% file: yields_energy.tex
\documentclass[a4paper,fleqn,usenatbib]{mnras}
\pdfoutput=1
\usepackage{newtxtext,newtxmath}

\usepackage[T1]{fontenc}
\usepackage{ae,aecompl}

\usepackage{graphicx}	
\usepackage{amsmath}	
\usepackage{amssymb}	

\usepackage{longtable}
\usepackage{pdflscape}
\usepackage{color}
\usepackage{subcaption}
\graphicspath{{figures/}}

\newcommand{\Msun}{\ensuremath{\mathrm{M}_\odot}} 

\title[Hypernovae Nucleosynthesis]{Nucleosynthesis in Primordial Hypernovae}

\author[J.J.\ Grimmett et al.]{
J.J. Grimmett,$^{1}$\thanks{E-mail: james.grimmett@monash.edu}
Alexander Heger,$^{1,2}$
Amanda I. Karakas,$^{1}$
Bernhard M\"uller$^{1,3}$
\\
$^{1}$Monash Centre for Astrophysics, School of Physics and Astronomy, 19 Rainforest Walk, Monash University, VIC 3800, Australia\\
$^{2}$Tsung-Dao Lee Institute, Shanghai 200240, P. R. China\\
$^{3}$Astrophysics Research Centre, School of Mathematics and Physics, Queen's University Belfast, BT7 1NN, Belfast, Northern Ireland
}

\date{Accepted XXX. Received YYY; in original form ZZZ}

\pubyear{2017}

\begin{document}

\label{firstpage}
\pagerange{\pageref{firstpage}--\pageref{lastpage}}
\maketitle

\begin{abstract}
We investigate the relationship between explosion energy and nucleosynthesis in Population III supernovae and provide nucleosynthetic results for the explosions of stars with progenitor masses of $15\,\Msun$, $20\,\Msun$, $30\,\Msun$, $40\,\Msun$, $60\,\Msun$, and $80\,\Msun$, and explosion energies between approximately $10^{50}$~erg and $10^{53}$~erg. We find that the typical abundance pattern observed in metal-poor stars are best matched by supernovae with progenitor mass in the range $15\,\Msun$ -- $30\,\Msun$, and explosion energy of $\sim(5 - 10) \times 10^{51}$~erg. In these models, a reverse shock caused by jumps in density between shells of different composition serves to decrease synthesis of chromium and manganese, which is favourable to matching the observed abundances in metal-poor stars.
Spherically symmetric explosions of our models with progenitor mass $\geq 40\,\Msun$ do not provide yields that are compatible with the iron-peak abundances that are typically observed in metal-poor stars, however, by approximating the yields that we might expect from these models in highly aspherical explosions, we find indications that explosions of stars $40\,\Msun$ -- $80\,\Msun$ with bipolar jets may be good candidates for the enrichment sources of metal-poor stars with enhanced carbon abundances.
\end{abstract}




\section{Introduction}

The nature of the first stars to exist in the Universe, and the role that they play in early chemical enrichment has persisted as a topic of interest for many years, yet still remains a mystery \citep{ezer_1971,carlberg_1981,silk_1983,audouze_1995,nakamura_1999,madau_2001,heger_2002,bromm_2003,wada_2003,tumlinson_2004,umeda_2005,greif_2007,tominaga_2007,heger_2010,nomoto_2013,frebel_2015,greif_2015}. Recent investigations suggest that the initial mass function (IMF) of the first stars, also known as Population III (Pop III) stars, may have been top heavy, dominated by massive stars ($>10\,\Msun$) \citep{susa_2014,hirano_2014,hosokawa_2016}. Furthermore, stars of zero metallicity likely experienced negligible mass loss throughout their evolution due to a lack of stellar winds \citep{kudritzki_2000,nugis_2000}, and may retain their initial mass up until their final fate. Perhaps the only property of these stars that can be asserted with reasonable confidence is their composition, which must reflect that of the primordial Universe from which they formed (i.e., Z=0, though see \citealt{jedamzik_2000}). Those Pop III stars that exploded as supernovae enriched the early Universe with heavy elements. Key questions that remain unanswered include whether the explosions of the first stars leave a unique chemical imprint on the Universe? If so, are we able to detect this imprint, and what can we learn about the first stars from it?\\

The next generation of stars to form immediately following the explosion of the first stars are thought to contain the ashes of one, or only a very few, supernovae \citep{audouze_1995}. This has two implications: (i) these early second generation stars would be metal-poor to a large degree, and (ii) the abundance profile of early second generation stars should contain a clear fingerprint of the chemical yield from either one, or a small number of the first supernovae. The chemical yields of a supernova reveals information about the explosion mechanism, which in turn reveals details of the progenitor star (e.g., mass, rotation rate, binarity), so these early second generation stars may contain our best clues to inferring the nature of the first stars \citep[e.g.,][]{frebel_2015}\\

Considerable progress has already been made in establishing the connection between the first generation of supernovae and very early second generation stars. A number of groups have made use of recently observed chemical abundances in metal-poor stars to compare to yields from supernovae calculations and galactic chemical evolution (GCE) models \citep{timmes_1995,chieffi_2002,umeda_2002,kobayashi_2006,tominaga_2007,joggerst_2009,joggerst_2010,heger_2010,nomoto_2013}. While GCE studies typically include multiple chemical enrichment sources (Type II, Type Ia, AGB stars, etc.) to model chemical evolution through several stellar generations, in the earliest epochs the yields from core collapse supernovae should dominate chemical abundances due to the short lifetimes of the massive progenitor stars. \\

Early GCE studies by \citet{timmes_1995} revealed that the chemical yields available from core collapse models at the time did not provide a good match to the observed abundances of iron peak elements in the most metal poor stars; [(Mn,Co,Zn)/Fe] were all underproduced.
Observations of the most metal poor stars have revealed a particularly interesting trend in the ratios of the iron-peak elements Cr, Mn, Co, and Zn. In general, for decreasing [Fe/H], [(Cr,Mn)/Fe] are seen to decrease whereas [(Co,Zn)/Fe] increases \citep{mcwilliam_1995b,ryan_1996,cayrel_2004}. Here the square brackets indicate the logarithm of the number ratio observed in the star relative to the solar value, i.e., [A/B] = log(N$_\mathrm{A}$/N$_\mathrm{B}$) - log(N$_\mathrm{A}$/N$_\mathrm{B}$)$_\odot$, where N refers to the number of moles of each respective element \citep{beers_2005}.\\
Uncertainties in both the mechanics of the supernova explosion, and in the evolution of the progenitor star, leaves some room for a parameter space in the supernova models. The ejected masses of iron peak elements are sensitive to the explosion parameters used in most one or two-dimensional supernova models, such as the mass cut, neutron excess profile, and mixing/fallback processes. A fine tuning of these parameters has been shown to give nucleosynthetic results which better reproduce this general trend in decreased [(Cr,Mn)/Fe] and increased [(Co,Zn)/Fe] \citep[e.g.,][]{nakamura_1999,umeda_2002,umeda_2005,heger_2010}.\\

Finding a combination of model parameters which provide nucleosynthetic results to \textit{quantitively} match the extreme values observed in the most metal poor stars however, has proven difficult. For example, a mass cut deep enough to produce sufficient [Zn/Fe] also requires a mixing and fallback mechanism to reduce the total mass of ejected iron, such that the intermediate mass element ratios [X/Fe] are not reduced too significantly. Adjusting the neutron excess profile has also been shown to be effective in altering the ratios of iron-peak elements \citep{nakamura_1999,umeda_2005}, but, again, requires some fine tuning of other parameters to maintain reasonable ratios of intermediate mass elements. No combination of these parameters for a spherically symmetric supernovae of typical explosion energy (i.e., of order $1\times10^{51}$~erg) has been shown to quantitatively reproduce the observed ratios in the most metal poor stars. \\

The discovery of hypernovae\footnote{The term "hypernova" was originally applied to the bright supernovae associated with GRB events \citep{paczynski_1998}. Over time, it is become more commonly used to describe highly energetic supernovae. In this study, we refer to supernovae with explosion energy of order 10~B as hypernovae.}, that is, supernovae with explosion energies $\sim 10$~B (1 Bethe (B) = 10$^{51}$ erg) \citep[e.g.,][]{galama_1998,iwamoto_1998,iwamoto_2000,matheson_2003,woosley_2006}, has motivated investigations into the relationship between supernova explosion energy and chemical yield \citep{nakamura_2001,umeda_2002,nomoto_2006}. This has provided promising results. 
\citet{nakamura_2001} and \citet{umeda_2005} have found that increasing the explosion energy by an order of magnitude from the canonical value of 1~B causes the silicon burning regions to extend outward in mass, which results in a larger mass ratio between the complete and incomplete silicon burning products. This is similar to the effect of deepening the mass cut, and is favourable to matching the observed trend of decreasing [(Cr,Mn)/Fe] and increasing [(Co,Zn)/Fe] ratios for stars of decreasing [Fe/H]. We should mention that, while angular momentum transfer and magnetic fields are not followed in our models, it is likely that the extra source of explosion energy is provided by rotation and magnetic fields in the progenitor star, which can also induce asphericity in explosions \citep[e.g.,][]{leblanc_1970,macfadyen_1999,burrows_2007,winteler_2012,mosta_2014,obergaulinger_2017}.
The symmetry of the explosion has also been shown to effect the ratios of ejected elements. Specifically, explosions with bipolar jets have been shown to produce the iron-peak, and some intermediate-mass (C, Sc, Ti) elements in similar ratios to those observed in EMP/CEMP (extremely metal poor/carbon enhanced metal poor) stars \citep{maeda_2003,tominaga_2007,tominaga_2009}. Jets have also been implicated as a potential source for r-process nucleosynthesis \citep{winteler_2012,nakamura_2015,mosta_2017,nishimura_2017}. Due to the stronger shock at the poles, these models provide a natural fallback effect, as much of the equatorial material is not provided with sufficient energy to be ejected. \\

The unique explosive yields that hypernovae are calculated to eject into the Universe has motivated their inclusion into GCE calculations \citep{kobayashi_2006,nomoto_2006,nomoto_2013}. These chemical evolution models typically include a free parameter to set the fraction of hypernovae to be included in the calculation. The explosion energy for the hypernova models is typically 10~B, though this prescribed value of explosion energy is somewhat arbitrary. High energy supernovae explosions are observed with a wide range of explosion energies. For example, the observations of SN1990E, SN1992am \citep{hamuy_2003}, SN1997ef \citep{iwamoto_2000}, SN1998bw \citep{woosley_1999} are well reproduced with models of explosion energies, 3.4~B, 5.5~B, 8~B, and 20~B, respectively. As it stands, it is not clear exactly where the boundary between supernova and hypernova lies in terms of nucleosynthesis, or whether a clear boundary even exists at all.\\

Our aim here is to investigate precisely how the ratios of certain heavy elements change with explosion energy. In Section \ref{sec:method}, we describe our initial models and computational procedure. In Section \ref{sec:results} we provide new results for the nucleosynthetic yields from the explosions of stars with initially zero metallicity and progenitor masses $15\,\Msun$, $20\,\Msun$, $30\,\Msun$, $40\,\Msun$, $60\,\Msun$, and $80\,\Msun$, for a fine grid of explosion energies between approximately $10^{50}$ erg and $10^{53}$ erg. In Section \ref{sec:jets} we provide a crude estimate of nucleosynthetic yields for explosions with bipolar jets. Finally, in Section \ref{sec:discussion}, we provide a discussion of our results including the potential application of our results to GCE models.\\

\section{Method}\label{sec:method}
\subsection{\textsc{KEPLER} Code}
We use the \textsc{Kepler} \citep{weaver_1978} stellar evolution code for all hydrodynamic and nucleosynthesis calculations. \textsc{Kepler} is a one-dimensional implicit hydrodynamics code suitable to applications in stellar evolution and explosive nucleosynthesis. \textsc{Kepler} makes use of an adaptive nuclear reaction network to follow nucleosynthesis, whereby the number of isotopes in the reaction network is automatically adjusted to accomodate the thermodynamic conditions of the environment \citep{rauscher_2002}. As we are performing our calculations in one dimension, we are assuming spherical symmetry in our models. We neglect the effects of rotation and magnetic fields.

\subsection{Initial Models and Procedure}\label{ssec:initialmodels}
We begin with several stellar models of masses $15\,\Msun$, $20\,\Msun$, $30\,\Msun$, $40\,\Msun$, $60\,\Msun$, and $80\,\Msun$, with corresponding helium core masses $3.7\,\Msun$, $5.58\,\Msun$, $9.95\,\Msun$, $15.29\,\Msun$, $23.9\,\Msun$, and $31.39\,\Msun$. These models are at the point of core collapse when our calculations begin. The evolution prior to this stage has been followed by \citet{heger_2010}, from an initial composition reflecting that of unpolluted big bang material. Rotation and mass loss were neglected in the evolution of these models. Although it may be that mass loss can be neglected for non-rotating zero metallicity stars, rotation is likely to have played a role in the evolution of the first stars \citep{stacy_2011}. Rotation enhances mass loss during the evolution, induces increased mixing inside the model, causes changes to the internal structure prior to collapse, and as a result has been shown to also affect the presupernova nucleosynthetic evolution by increasing synthesis of nitrogen and also enhancing s-process production at low metallicity \citep{meynet_2006,hirschi_2007,pignatari_2008,maeder_2015,frischknecht_2016}. Rotation, when very rapid, also results in asymmetries in the explosion, and likely also provides the energy source for explosions with hypernova-like energy \citep{macfadyen_1999, macfadyen_2001,burrows_2007}. Rotation will therefore have a significant effect on the nucleosynthetic results \citep[e.g.,][]{tominaga_2009}. We will reserve an investigation including the effects of rotation during collapse for a future study. Here we are primarily interested in the effect of explosion energy on nucleosynthesis without considering the mechanism by which a collapsing star might be able to achieve such large ($\gg 1$~B) explosion energies. Explosion energies of this magnitude are beyond the reach of the neutrino-driven mechanism, and as such, we also neglect neutrino transport in our models. We assume that whatever the hypernova mechanism is in reality, ejecta are immediately heated and ejected by a powerful shock, and not exposed to any significant neutrino flux from the cooling neutron star as in the neutrino-driven mechanism, which would otherwise affect nucleosynthesis \citep[e.g.,][]{buras_2006,marek_2009,muller_2012a,frohlich_2014}. Although there may be another ejecta component that is more strongly neutrino-processed or has undergone stronger deleptonisation (e.g., from a proto-neutron star or collapsar disk wind), nucleosynthesis by explosive burning will always contribute and can justifiably be investigated using our piston-driven models.\\

We use the progenitor models described above to simulate core collapse by means of an artificial piston, as in \citet{heger_2010}. By this method, material near the iron core is moved briefly inward at approximately $1/4$ local free-fall acceleration, then launched outward, also at a constant fraction of the local gravitational acceleration, where the factor is chosen in order to achieve the desired explosion energy, as is described in \citet{rauscher_2002}. A crucial parameter is the piston location, which determines the initial mass cut. This is the boundary between the central remnant and the ejected mass, before fallback is considered. Lacking a fundamental understanding of the explosion engine, an appropriate location for the piston remains uncertain. Some constraints can be placed on the piston location by the resulting ejected chemical abundances and central remnant masses; if the piston is too deep, a successful explosion will overproduce neutron-rich species in the iron group, and if the piston is too shallow, the typical neutron star mass will be too large to match with observed values \citep{woosley_2007}. Following \citet{woosley_2003} and \citet{heger_2010}, we take the piston location to be at a density discontinuity near the base of the oxygen shell in the pre-supernova model, which generally occurs around S/N$_\mathrm{A}$k$_\mathrm{b}$ $\simeq 4$ (except for our $40\,\Msun$ model, where the density discontinuity is closer to S/N$_\mathrm{A}$k$_\mathrm{b}$ $\simeq 5$). Here  S/N$_\mathrm{A}$k$_\mathrm{b}$ is the entropy per baryon.\\

The parametrised approach that we implement is necessary for large scale supernovae nucleosynthesis studies. Self-consistent multidimensional supernova models are too computationally expensive for use in studying even a modest parameter space, and besides, are not yet robust enough for such a study. Admittedly, the parametrisation in our models leaves significant room for uncertainties in results. The method to initiate an explosion, piston location, and details of the progenitor model all come with corresponding uncertainties, and on top of this, the nuclear reaction rates relevant to explosive silicon burning, the dominant nucleosynthesis process in our models, are as yet not entirely well constrained \citep[e.g.,][]{woosley_1973,hix_1999,sonzogni_2000,descouvemont_2006,simon_2013,rauscher_2013}. Nevertheless, these parametrised explosions will be used for the foreseeable future in supernovae nucleosynthesis studies, and it is important that we properly understand how sensitive our results are to each parameter. Here we are specifically interested in the relationship between explosion energy and chemical yield, and we point the reader toward several other studies which investigate the effects of varying other model parameters \citep{aufderheide_1991,nakamura_1999,nakamura_2001,umeda_2002,umeda_2005,tominaga_2007,young_2007,joggerst_2010,heger_2010,fryer_2018}. The uncertainties and implications of the piston-driven explosion will be discussed further in Section \ref{sec:discussion}.\\

For each progenitor model we simulate explosions with a minimum of 50 different explosion energies between 0.05 B and 100 B. This grid of explosion energies is not equally spaced, and we increase the resolution for regimes where we see interesting transitions in physical behaviour. We calculate our final yields with an adaptive nuclear reaction network, which can include up to $\sim$5000 isotopes. \textsc{Kepler} does have capacity to include an approximation for mixing effects, whereby composition is averaged across mass elements prior to ejection. As our primary objective is to investigate the effect of explosion energy on chemical yield, we do not include mixing in the majority of our analysis, so as not to confuse the effects of mixing with those of explosion energy. We explicitly state where mixing is included in particular results, and note for all models with large explosion energy (most of our models), all outer layers are ejected and mixing is then irrelevant.\\

\section{Nucleosynthetic Results}\label{sec:results}
\onecolumn
\begin{figure}
	\includegraphics[width=\columnwidth]{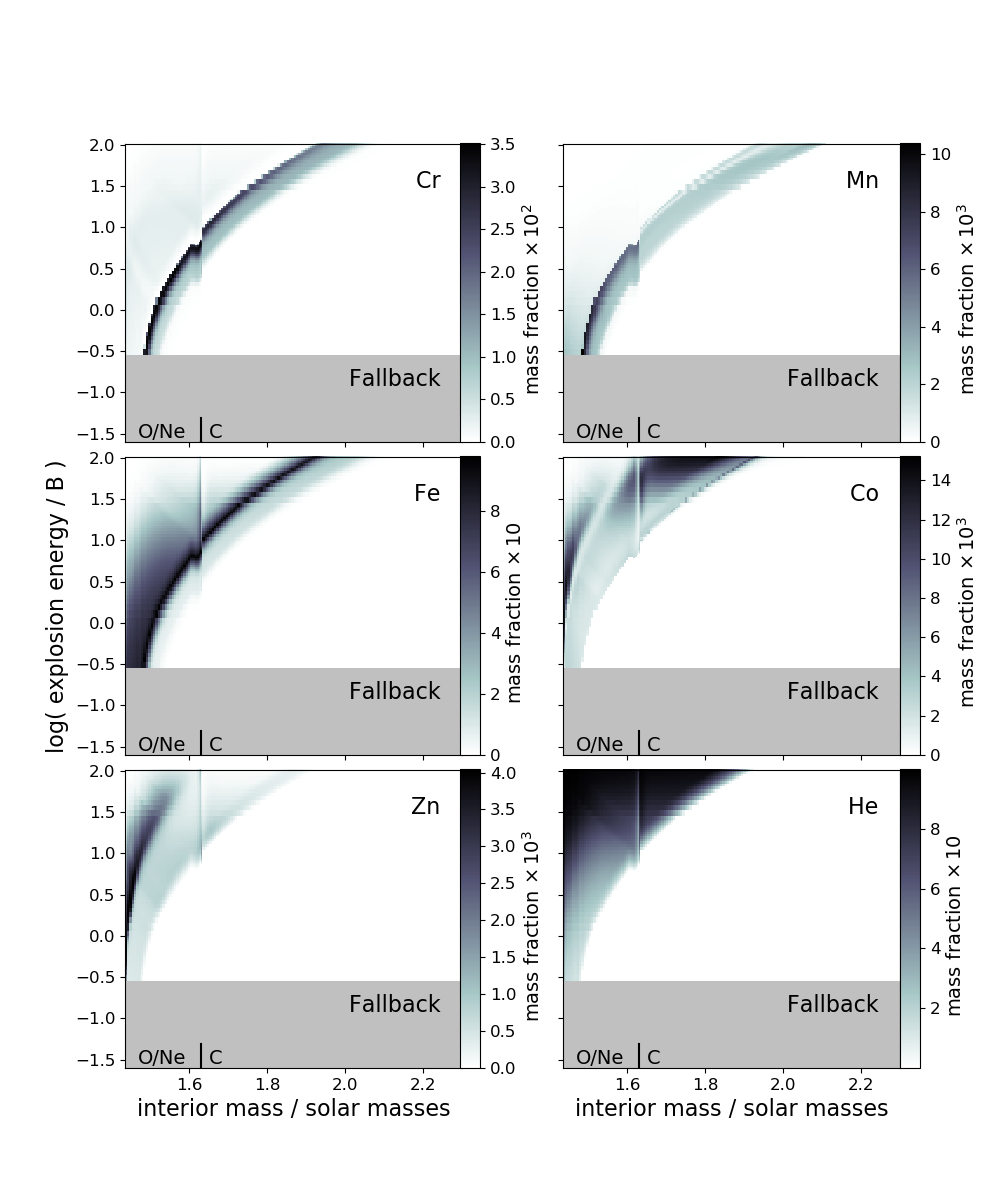}
    \caption{Mass fractions after decay of chromium, manganese, iron, cobalt, zinc, and helium (from photodisintegration) as a function of mass coordinate and explosion energy for the $15\,\Msun$ progenitor. The grey shaded region indicates where mass has fallen back onto the central remnant for low explosion energy. The short lines on the bottom of each plot mark the boundaries between shells in the progenitor model, with the label indicating which element is the primary fuel for burning in each shell.}
    \label{fig:z15_totalnucleo}
\end{figure}

\begin{figure}
	\includegraphics[width=\columnwidth]{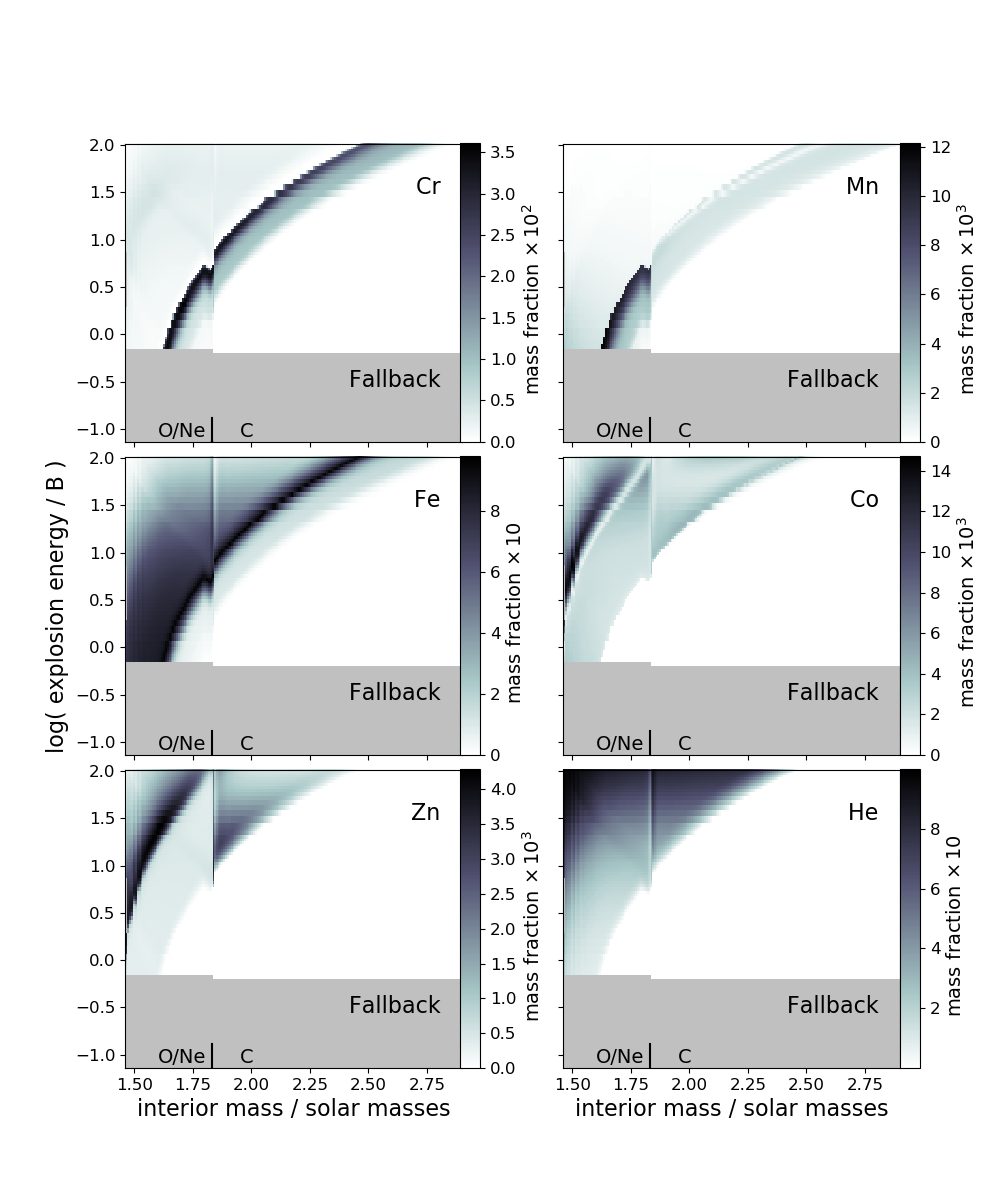}
    \caption{Same as Figure \ref{fig:z15_totalnucleo}, but for the $20\,\Msun$ model.}
    \label{fig:z20_totalnucleo}
\end{figure}

\begin{figure}
	\includegraphics[width=\columnwidth]{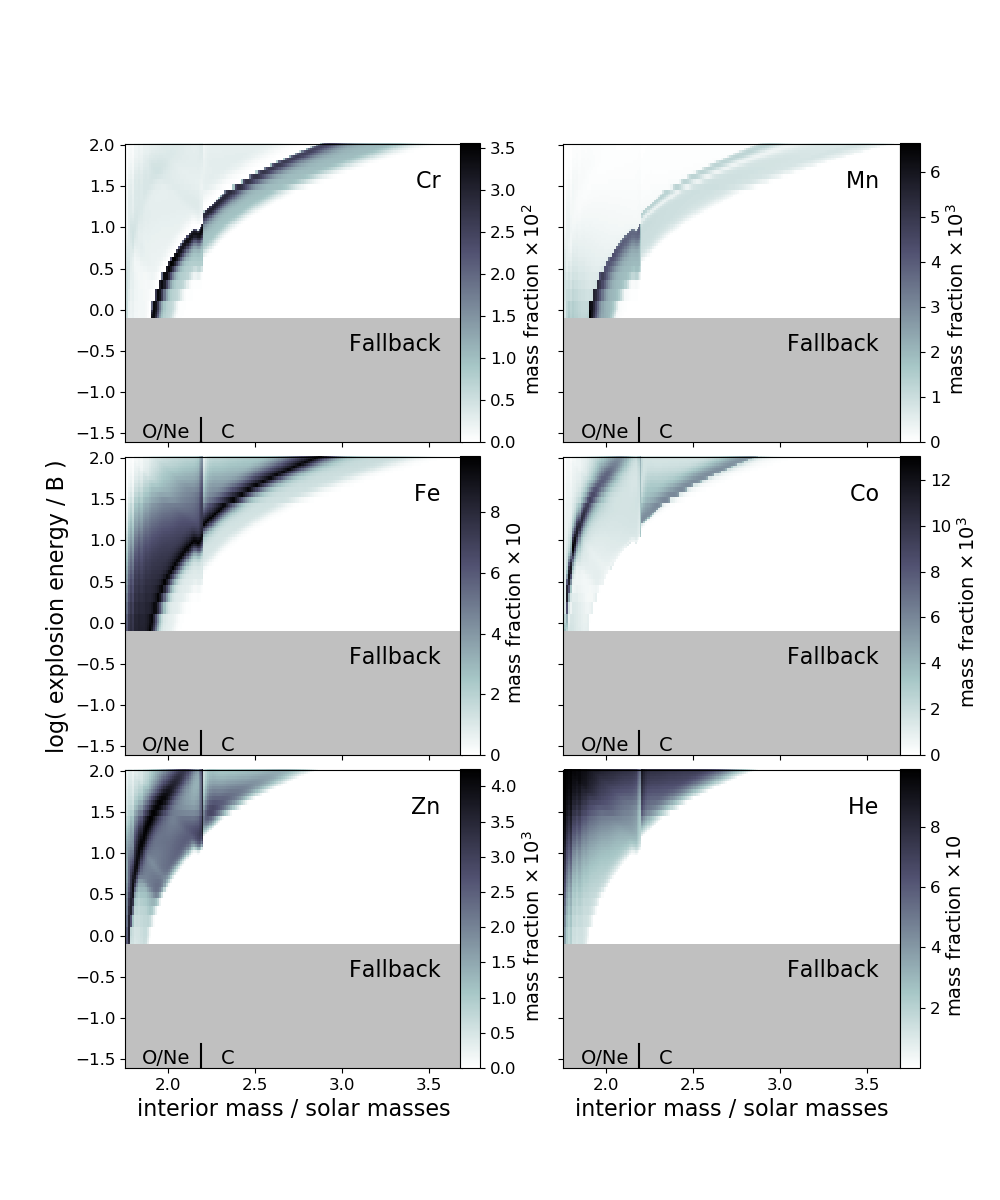}
    \caption{Same as Figure \ref{fig:z15_totalnucleo}, but for the $30\,\Msun$ model.}
    \label{fig:z30_totalnucleo}
\end{figure}

\begin{figure}
	\includegraphics[width=\columnwidth]{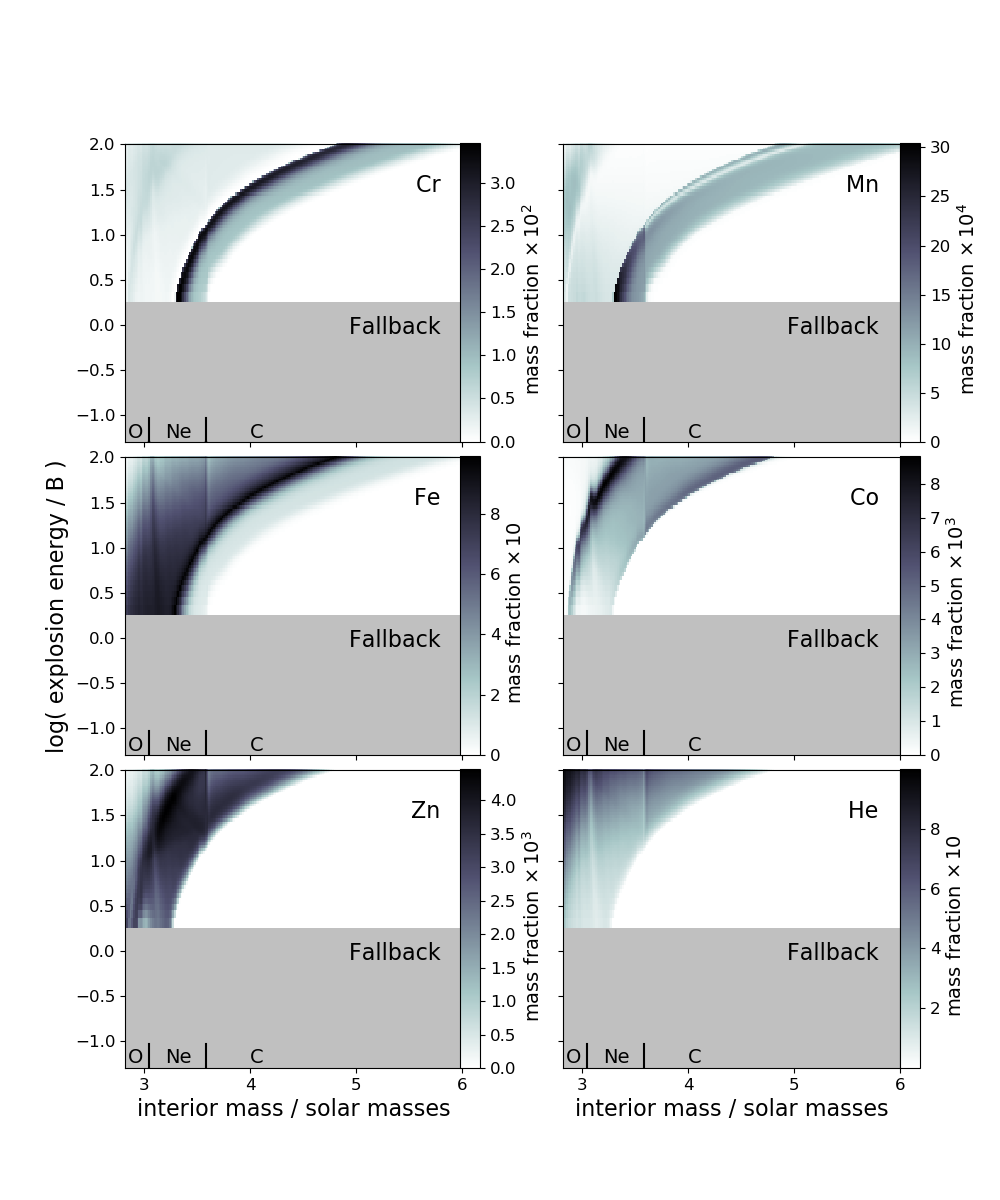}
    \caption{Same as Figure \ref{fig:z15_totalnucleo}, but for the $40\,\Msun$ model.}
    \label{fig:z41_totalnucleo}
\end{figure}

\begin{figure}
	\includegraphics[width=\columnwidth]{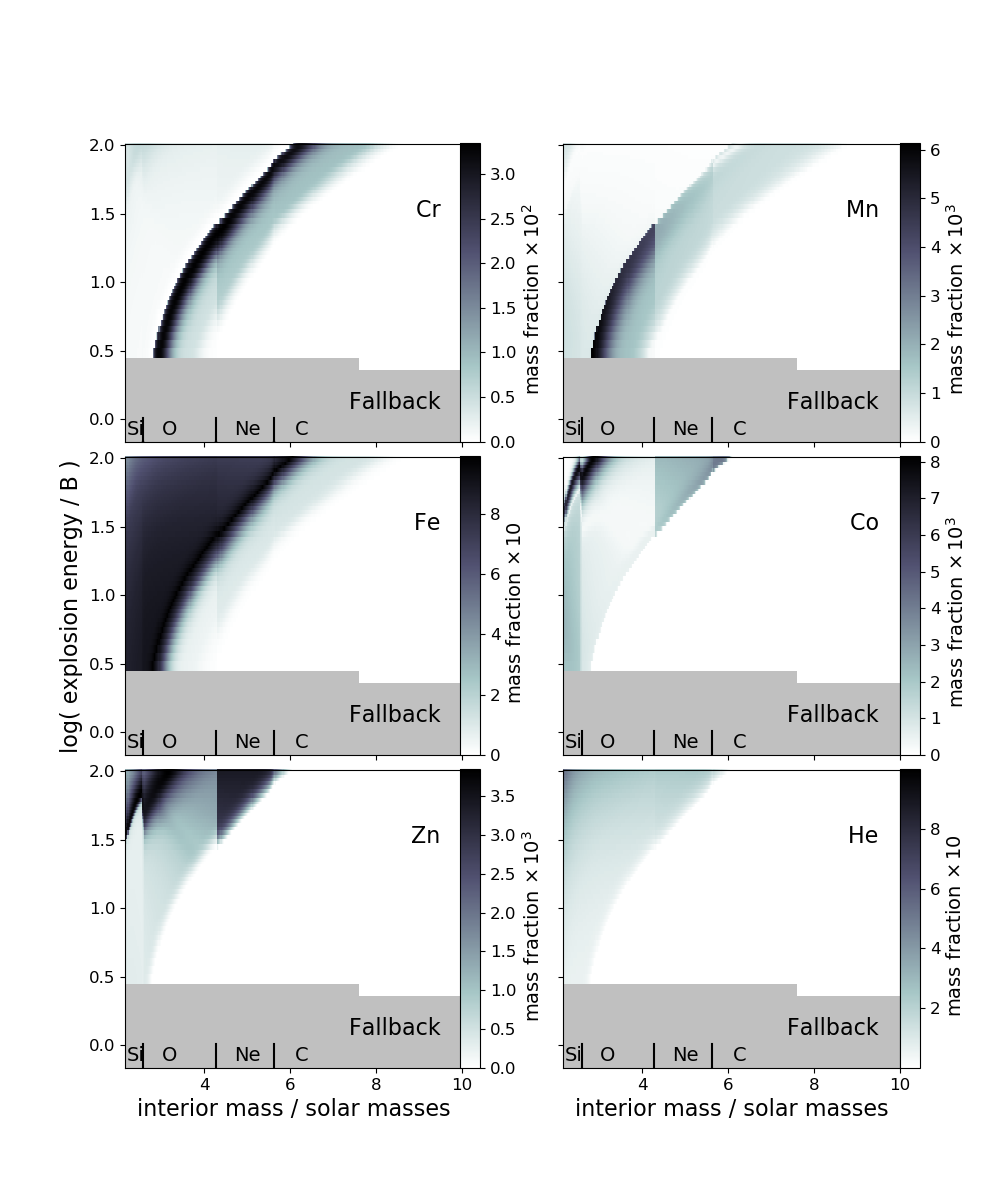}
    \caption{Same as Figure \ref{fig:z15_totalnucleo}, but for the $60\,\Msun$ model.}
    \label{fig:z60_totalnucleo}
\end{figure}

\begin{figure}
	\includegraphics[width=\columnwidth]{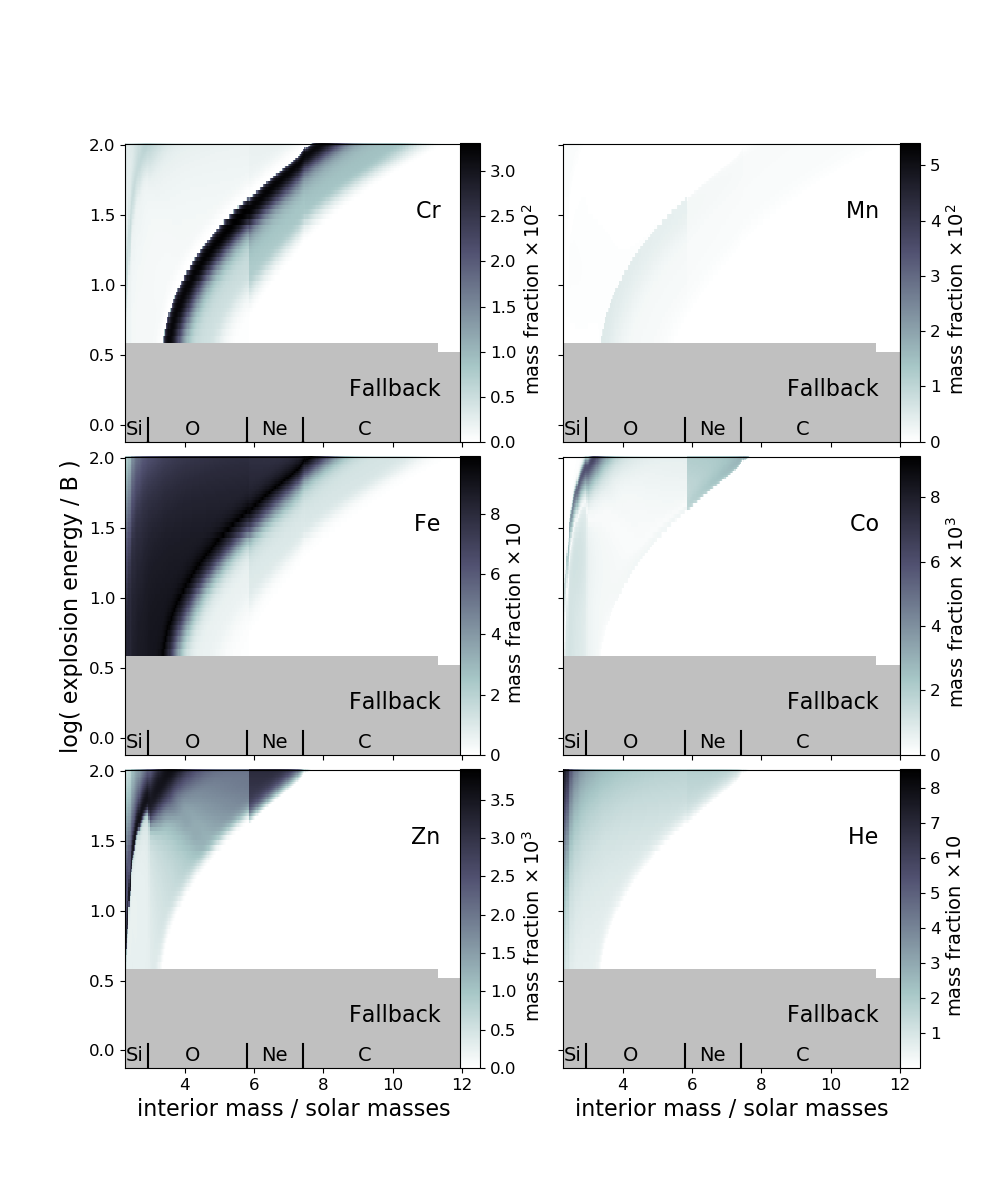}
    \caption{Same as Figure \ref{fig:z15_totalnucleo}, but for the $80\,\Msun$ model.}
    \label{fig:z80_totalnucleo}
\end{figure}
\twocolumn

We provide yields for all of our models as online downloadable tables. Here we first discuss some of the main features of our results in a general sense, before moving on to discuss the final abundances for some elements of particular interest. We focus on the iron-peak elements chromium, manganese, cobalt, and zinc, and briefly discuss key intermediate mass elements, namely, potassium, scandium, titanium, vanadium. In Figures \ref{fig:z15_totalnucleo} to \ref{fig:z80_totalnucleo}, we display our results for the nucleosynthesis of the iron-peak elements of interest, and also helium, as a function of explosion energy and interior mass coordinate. Chromium and manganese production reflect the region of incomplete silicon burning, whereas iron, cobalt, and zinc reflect complete silicon burning, and helium abundance reflects the varying degree of $\alpha$-rich freezeout.\\

We report that our results are in general agreement with those of previous investigations, in the sense that for increasing explosion energy, the entire silicon burning region extends outwards in mass, and the mass ratio between the complete silicon burning region and the incomplete silicon burning region increases. \citep[e.g.,][]{nakamura_1999,umeda_2002,nomoto_2006,tominaga_2007}.

We also find that in addition to the mass ratio between complete and incomplete silicon burning, effects such as the change in composition between shells, reverse shocks at shell interfaces, and fast expansion times of shocked material for high explosion energy, are all equally significant factors in determining the final [(Cr,Mn,Co,Zn)/Fe] abundances. These effects will be summarised before we move on to discussing the yields of specific elements in detail.\\

\subsection{Effects of Shell Composition on Nuclear Burning}\label{ssec:neutron_excess}
During complete silicon burning, material is photodisintegrated into protons, neutrons, and $\alpha$-particles. As the material expands and cools, these light particles recombine to heavier elements. The final composition is primarily set by the proton-to-neutron ratio and the timescale for expansion/cooling. Therefore, the final products of complete silicon burning depend sensitively on the temperature, density, and neutron excess of the fuel being burned \citep[e.g.,][]{woosley_1973,hix_1999}. These three properties differ between shells, and as such, we find that the products of nucleosynthesis also vary as silicon burning shifts outward for increasing explosion energy. That is, for the same burning process, the final products will differ depending on the location that the burning process occurs. For example, the decrease in neutron excess and faster expansion times at larger radius can be seen to particularly affect manganese production, shown in Figures \ref{fig:z15_totalnucleo} to \ref{fig:z80_totalnucleo}. The primary source of manganese, with just one stable isotope, $^{55}$Mn, is as the decay product of $^{55}$Co, which requires a neutron excess for synthesis. As seen in Figures \ref{fig:z15_totalnucleo} to \ref{fig:z80_totalnucleo}, the local manganese mass fraction is reduced for incomplete silicon burning at a larger mass coordinate. 
We also find that for explosion energies which increase the temperature in the second or third most inner shell (oxygen/neon burning or carbon burning shell, depending on the model) to levels appropriate for complete silicon burning, we achieve multiple sites of cobalt and zinc production, one in each shell where the temperature allows synthesis, for the given density and neutron excess. For example, this is particularly evident for zinc synthesis for the $20\,\Msun$ model (Figure \ref{fig:z20_totalnucleo}), or for cobalt and zinc synthesis in the $60\,\Msun$ model (Figure \ref{fig:z60_totalnucleo}). \\

\subsection{$\alpha$-Rich Freeze-Out}\label{ssec:alpharich}
A further consequence of the high temperatures resulting from large explosion energies is a fast expansion time of the shocked material. During complete silicon burning, all material is essentially photodisintegrated into $\alpha$-particles. For high temperature, low density, or a combination of both, the material may expand faster than the $\alpha$-particles are able to reassemble into heavy elements \citep[e.g.,][]{woosley_1973,hix_1999}. We see this in our models as an $\alpha$-rich freezeout, directly reflected by the final helium abundances shown in Figures \ref{fig:z15_totalnucleo} to \ref{fig:z80_totalnucleo}. For increasing explosion energy, a larger abundance of $\alpha$-particles remain. In fact, for our greatest explosion energies, the $\alpha$-particle abundance can begin to completely dominate the inner regions. This effect is much stronger in the lower mass models, which have lower inner densities and hence faster expansion times. We find that the extremely strong $\alpha$-rich freezeout, with X(He) $\gtrsim 0.6$, predominantly impacts the abundances of complete silicon burning products, reducing the local yield of cobalt and zinc for increasing explosion energy, and limiting iron synthesis to an increasingly small region. The effect propagates through to all abundance ratios when given in reference to iron, i.e., [X/Fe]. 

\subsection{Reverse Shock at Shell Interfaces}\label{ssec:rev_shock}
In Figures \ref{fig:z15_totalnucleo} - \ref{fig:z30_totalnucleo} ($15\,\Msun$, $20\,\Msun$, and $30\,\Msun$ models) we see that silicon burning extends steadily outward in mass with increasing explosion energy, unless the explosion energy is such that silicon burning is at the outermost edge of the oxygen burning shell. This shell boundary lies at $m\simeq1.65\,\Msun$, $m\simeq1.85\,\Msun$, and $m\simeq2.25\,\Msun$ for the $15\,\Msun$, $20\,\Msun$, and $30\,\Msun$ models, respectively. As silicon burning approaches this shell boundary for increasing explosion energy, we see that there is a particular explosion energy in each model for which incomplete silicon burning, reflected in Figures \ref{fig:z15_totalnucleo} - \ref{fig:z30_totalnucleo} by chromium and manganese synthesis, ceases abruptly inside the oxygen burning shell, and is instead replaced by complete silicon burning (reflected in Figures \ref{fig:z15_totalnucleo} - \ref{fig:z30_totalnucleo} by iron production). 

\begin{figure}
\centering
\includegraphics[width=0.9\columnwidth]{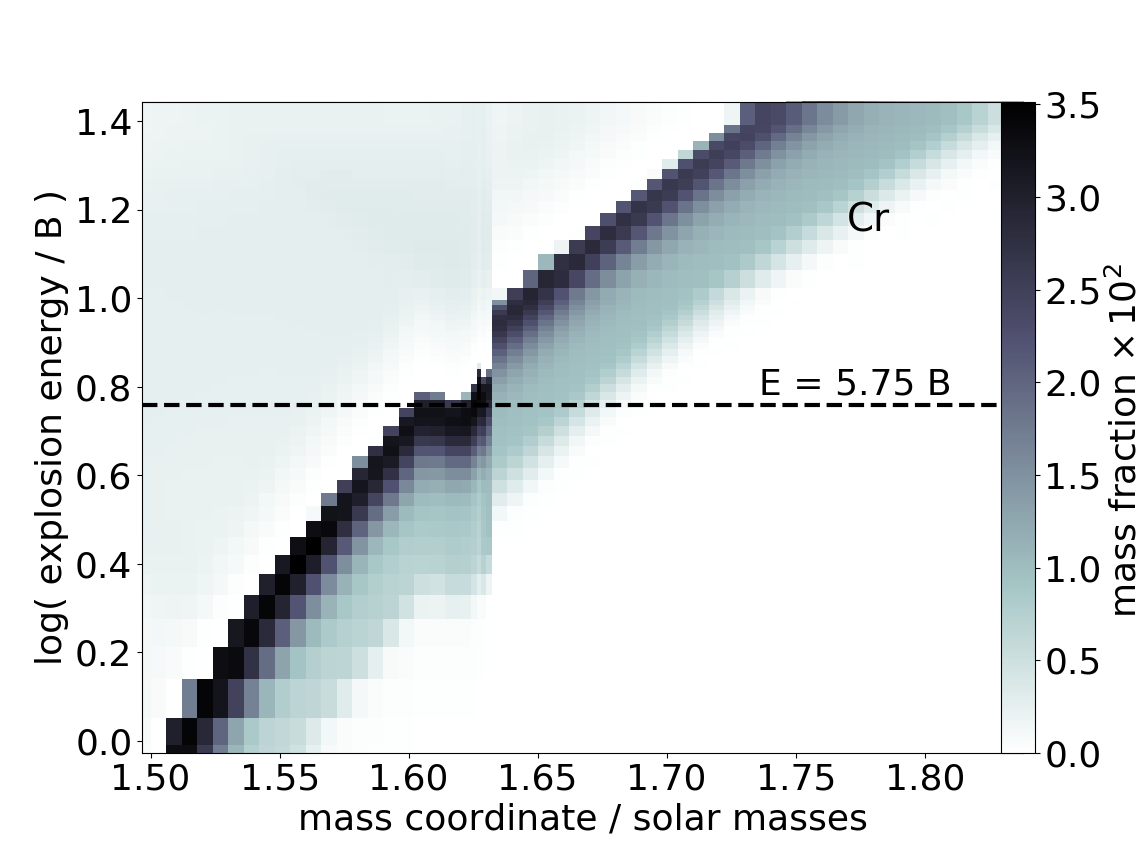}
\includegraphics[width=0.9\columnwidth]{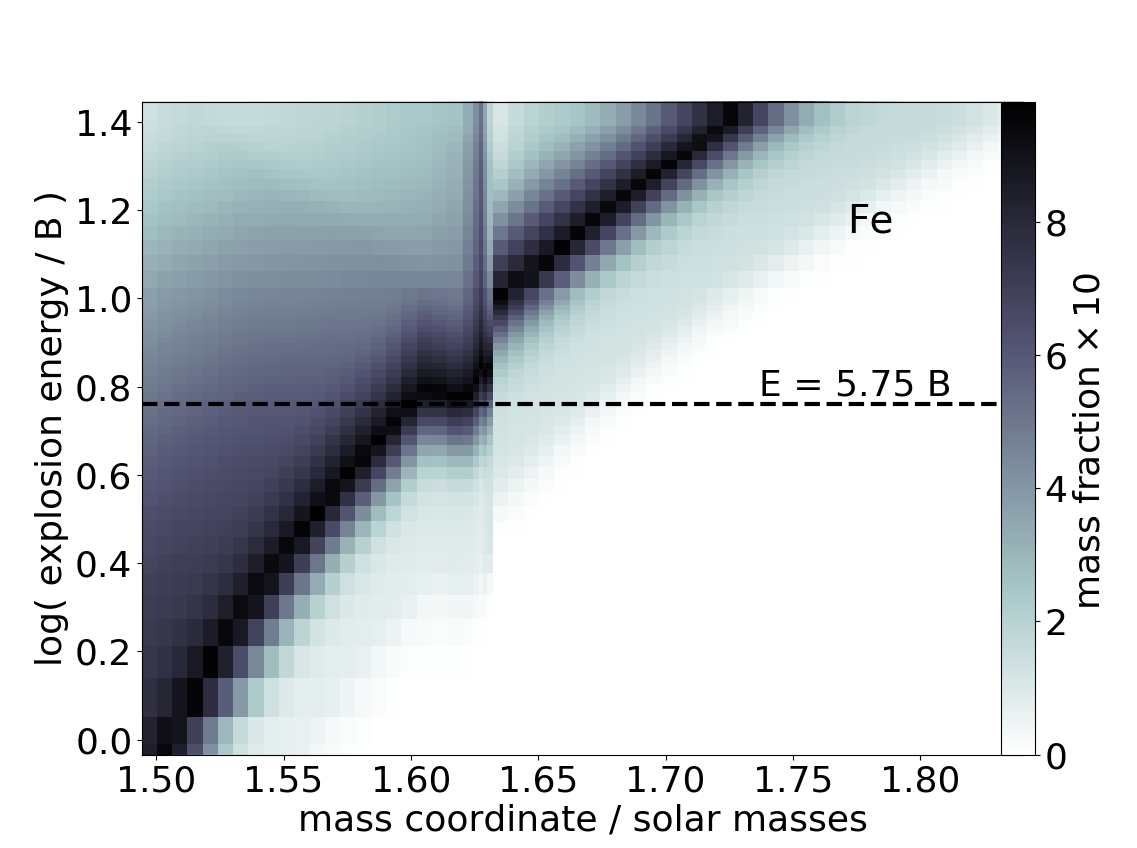}
\includegraphics[width=1.\columnwidth]{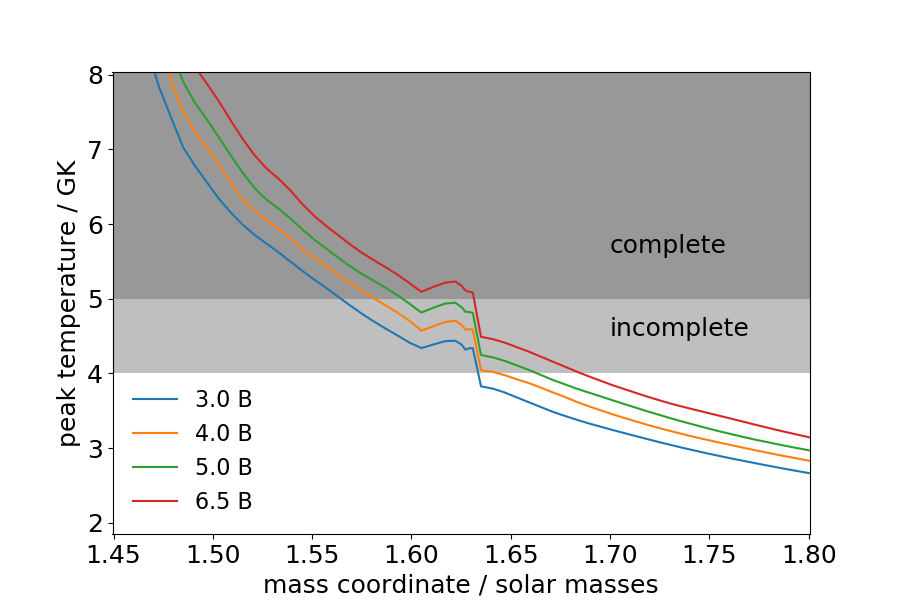}
\caption{In the upper and middle plots we show chromium and iron mass fractions, respectively, as a function of mass coordinate and explosion energy for the $15\,\Msun$ model, as in Figure \ref{fig:z15_totalnucleo}. The lower plot shows the peak temperature profiles for four models with explosion energy at which explosive silicon burning is affected by the reverse shock.}
\label{fig:nucleozooms}
\end{figure}

To explain the cause of this change in nucleosynthesis, we will use the $15\,\Msun$ model as an example for the remainder of this section, noting that we see the same effect in the $20\,\Msun$ and $30\,\Msun$ models, albeit at different explosion energies and shell boundary locations. The effect is best understood by examining Figure \ref{fig:nucleozooms}. The particular explosion energy for which incomplete silicon burning mostly ceases inside the oxygen burning shell is shown by the dashed line at 5.75 B in Figure \ref{fig:nucleozooms} (\textsl{top}). For explosion energy $1-2\,$B below this value, incomplete silicon burning occurs extensively against the inner edge of the boundary, and for greater explosion energy, incomplete silicon burning does not occur at all within the oxygen burning shell. Examining the peak temperature profile in Figure \ref{fig:nucleozooms} (\textsl{bottom}), we see a "bump" of increased temperature at the inner edge of the shell boundary (location shown by the dashed line), indicating that material in this region is subject to some source of extra heating. We understand the cause of the "bump" in the temperature profile against the outer edge of the oxygen burning shell as follows: as the initial supernova shock travels through the composition boundary, a reverse shock is launched as a consequence of the change in the sign of the gradient in $\rho r^3$ at this point. The quantity $\rho r^3$ is important for the shock velocity in supernovae. As is described by \citet{kompaneets_1960,klimishin_1982,bethe_1990,matzner_1999}, the shock velocity, $U$, can be approximated by

\begin{equation}
U \sim \left(\cfrac{1}{\rho r^3}\right)^{-\beta}.
\end{equation}\\

\begin{figure}
	\includegraphics[width=\columnwidth]{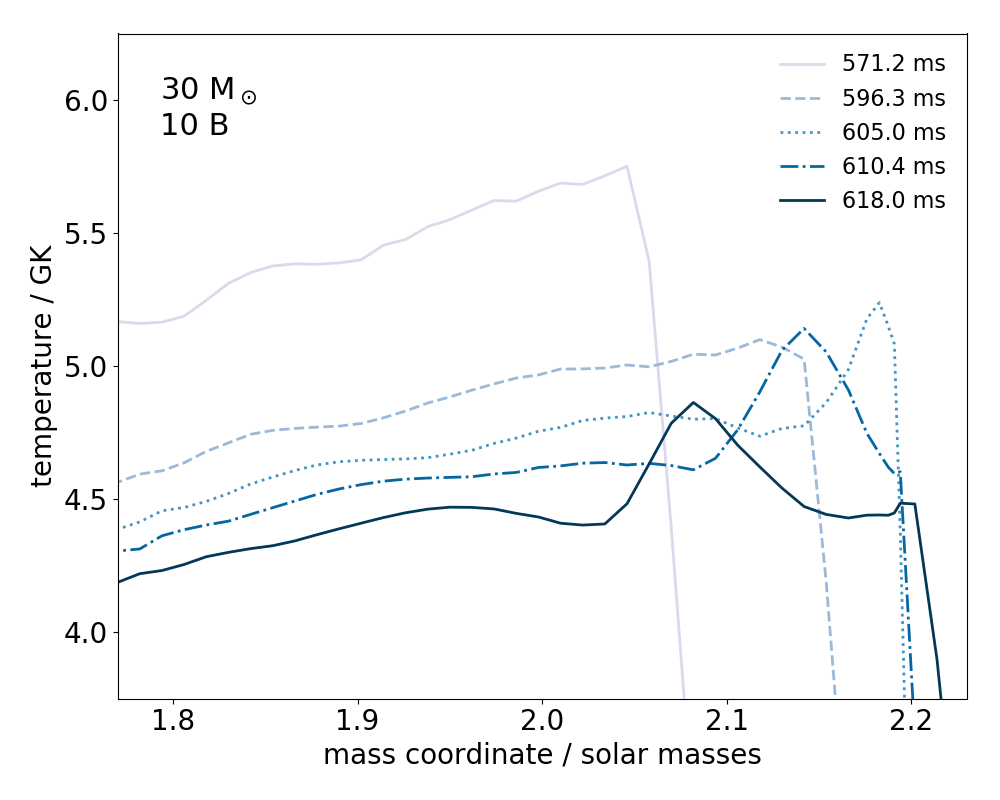}
    \caption{Temperature profile of the $30\,\Msun$ model with an explosion energy of $10\,$B. We show the profile for several time steps, revealing the reheating effect of the reverse shock, as the initial shock passes through the oxygen burning/carbon burning shell interface at $\sim 2.2\,\Msun$.}
    \label{fig:shockplot}
\end{figure}

\begin{figure}
    \centering
    \begin{subfigure}[b]{0.5\textwidth}
        \centering\captionsetup{width=.8\linewidth}
        \includegraphics[width=\textwidth]{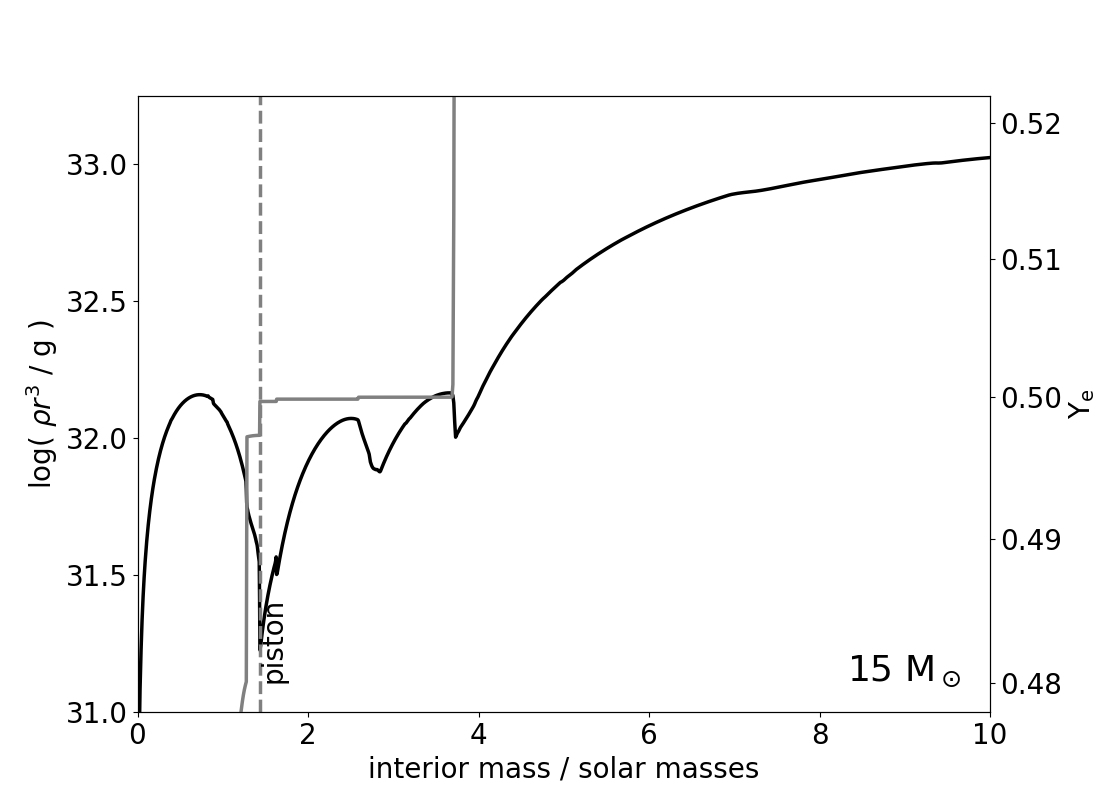}
		\caption{The $15\,\Msun$ model.}
		\label{fig:z15_shockconditions}
    \end{subfigure}%
    \begin{subfigure}[b]{0.5\textwidth}
        \centering\captionsetup{width=.8\linewidth}
        \includegraphics[width=\textwidth]{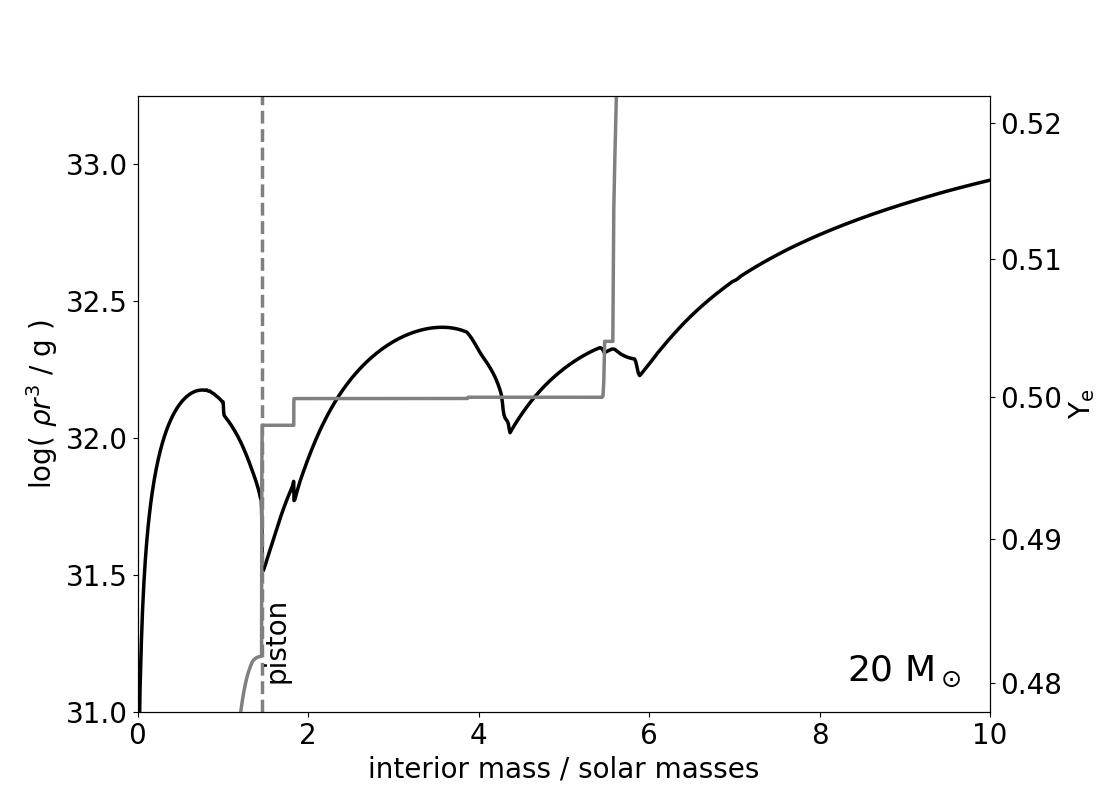}
		\caption{The $20\,\Msun$ model.}
		\label{fig:z20_shockconditions}
    \end{subfigure}
    \begin{subfigure}[b]{0.5\textwidth}
        \centering\captionsetup{width=.8\linewidth}
        \includegraphics[width=\textwidth]{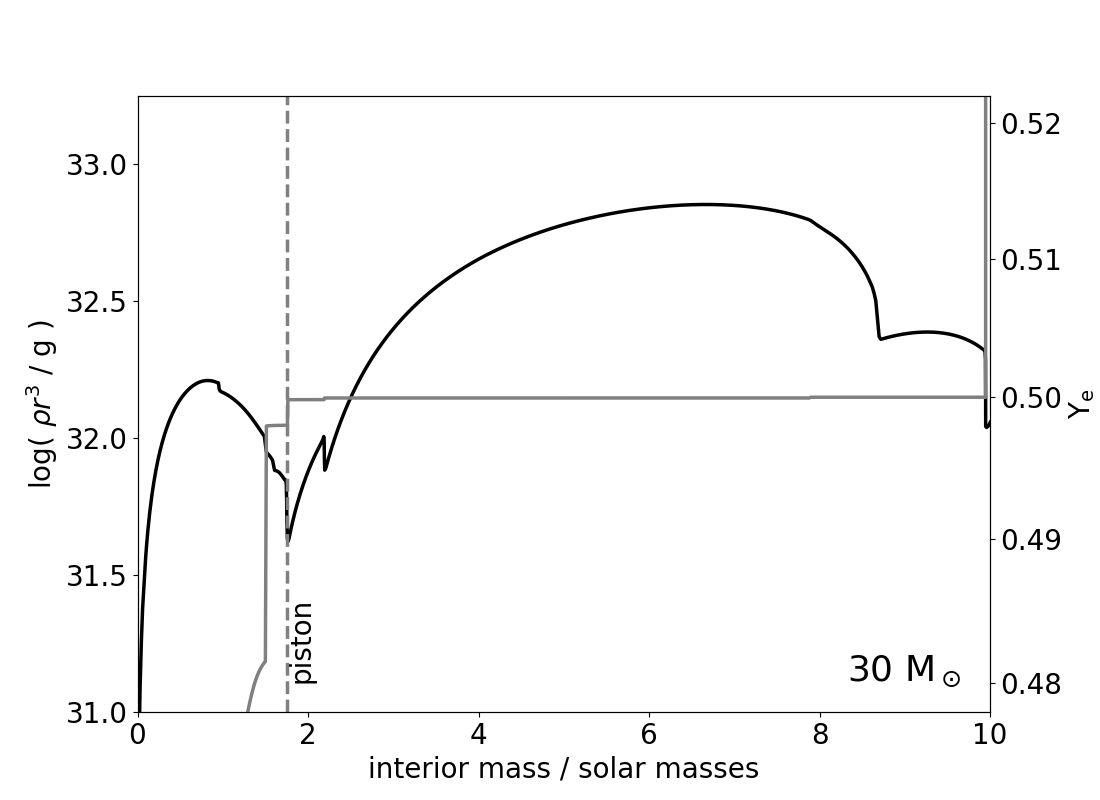}
		\caption{The $30\,\Msun$ model.}
		\label{fig:z30_shockconditions}
    \end{subfigure}%
    ~ 
    \begin{subfigure}[b]{0.5\textwidth}
        \centering\captionsetup{width=.8\linewidth}
        \includegraphics[width=\textwidth]{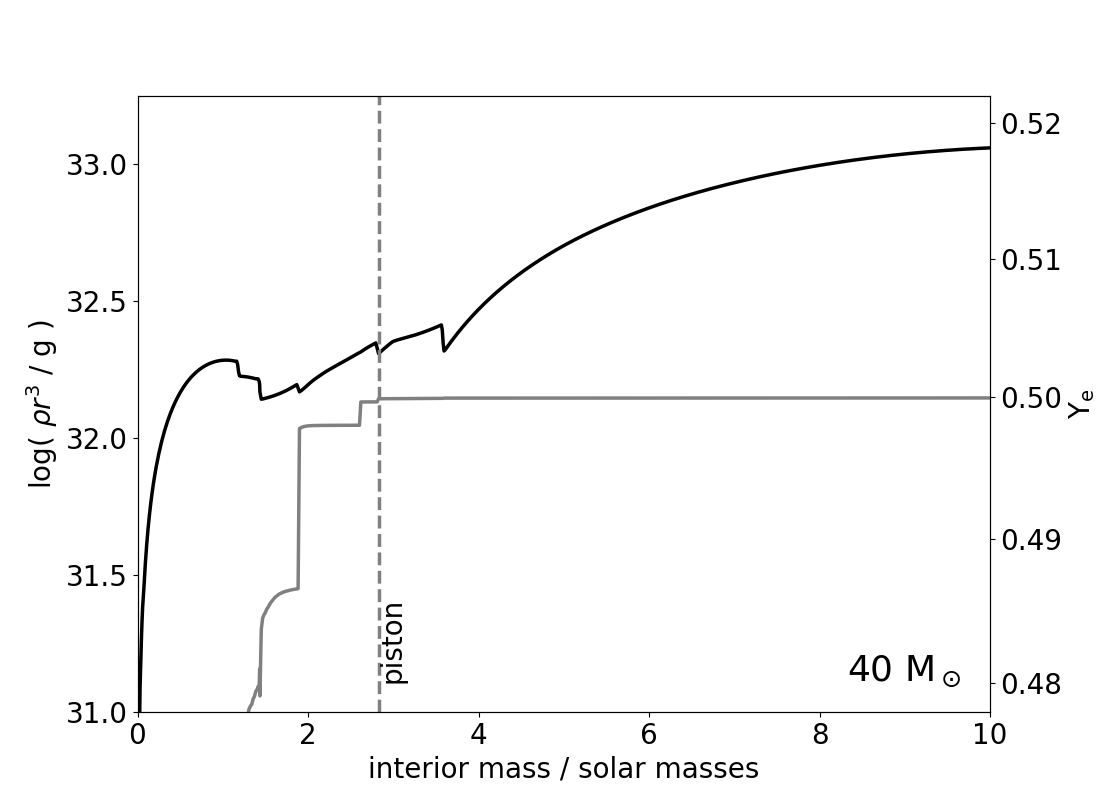}
		\caption{The $40\,\Msun$ model.}
		\label{fig:z41_shockconditions}
    \end{subfigure}
     \begin{subfigure}[b]{0.5\textwidth}
        \centering\captionsetup{width=.8\linewidth}
        \includegraphics[width=\textwidth]{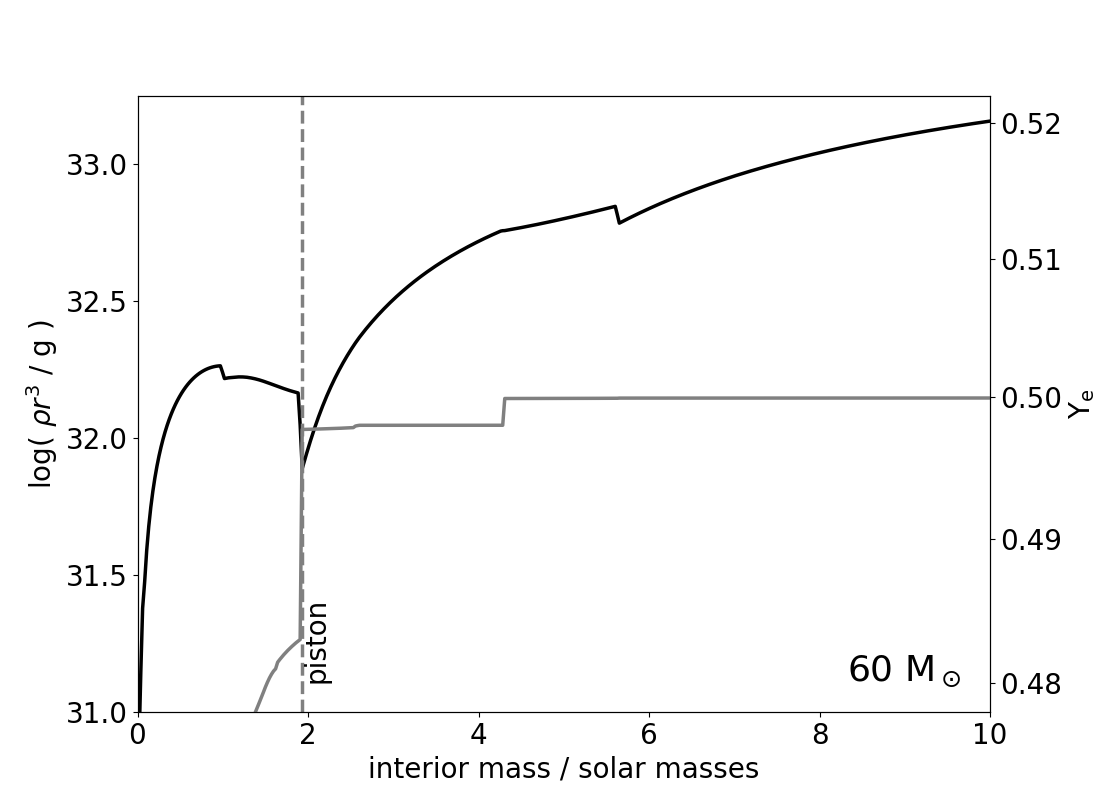}
		\caption{The $60\,\Msun$ model.}
		\label{fig:60_shockconditions}
    \end{subfigure}%
    ~ 
    \begin{subfigure}[b]{0.5\textwidth}
        \centering\captionsetup{width=.8\linewidth}
        \includegraphics[width=\textwidth]{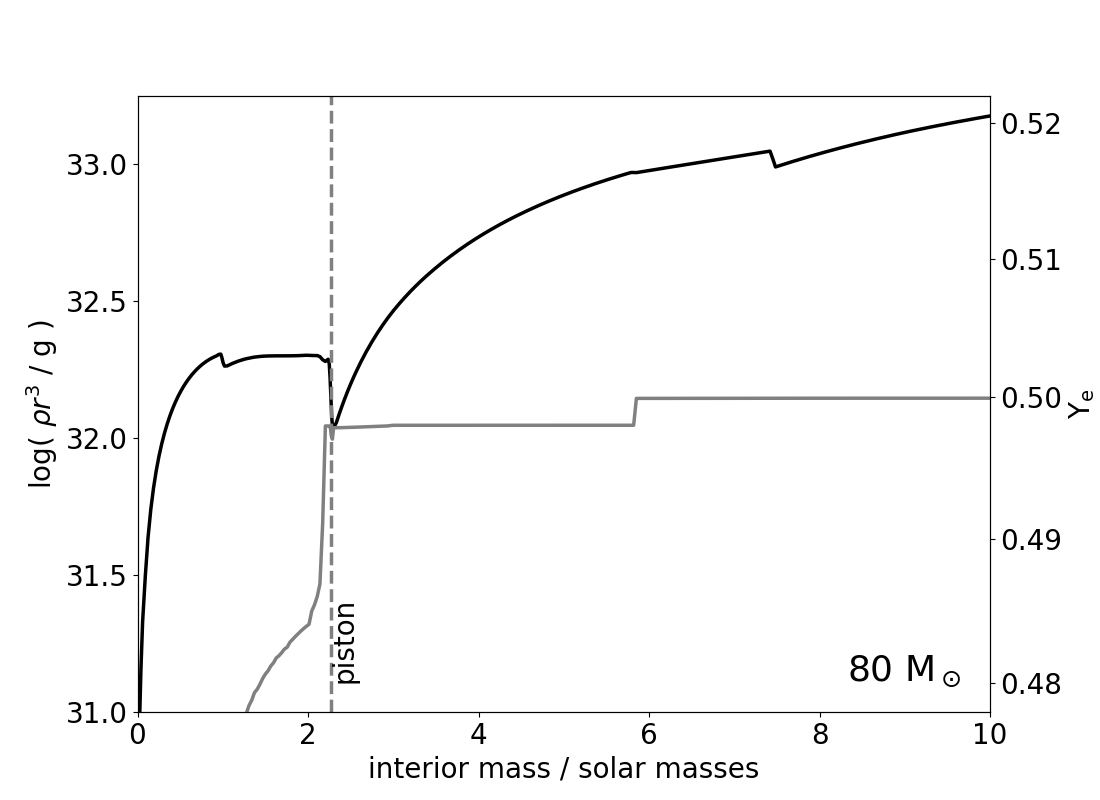}
		\caption{The $80\,\Msun$ model.}
		\label{fig:z80_shockconditions}
    \end{subfigure}
    \caption{The $\rho r^3$ profile (black line) and electron mole fraction, $Y_\mathrm{e}$ (grey line) for the inner $10\,\Msun$ of each of the $15\,\Msun$ -- $80\,\Msun$ progenitor models. The piston location is indicated by the dashed line. Note, Y$_\mathrm{e}=\frac{1}{2}(1-\eta)$, where $\eta$ is the neutron excess.}
\end{figure}

Here $\beta$ is parameter which varies with the properties of the material that the shock is propagating through, typically, $0.2 \lesssim \beta \lesssim 0.5$. Nevertheless, for the our purposes the important point is that the shock velocity is inversely proportional to $\rho r^3$, so when $\rho r^3$ is increasing the shock must decelerate, and when $\rho r^3$ is decreasing the shock accelerates. The inner $\rho r^3$ profile is shown for the $15\,\Msun$ model in Figure \ref{fig:z15_shockconditions}. We see that there is a reversal in the sign of the gradient of the $\rho r^3$ profile at the oxygen burning/carbon burning shell interface. Such a rapid change in $\rho r^3$, from a negative to positive gradient leads to a rapid slowing of the shock velocity. This abrupt deceleration results in an increased density at this point, as interior matter piles up against the slowed mass. This has a chain effect on the matter behind it, and so on, which propagates as an ingoing shock. \\
The reverse shock provides additional heating to the material immediately interior to the shell boundary, as can be seen in the temperature profile at various time intervals (Figure \ref{fig:shockplot}). This results in an approximately flat temperature profile for the region within $\sim0.03\,\Msun$ of the oxygen burning/carbon burning shell boundary, reflected by the extended region of chromium production (incomplete silicon burning) shown in Figure \ref{fig:nucleozooms} (\textsl{top}). 
For explosion energies greater than $\sim$ 5.75 B in this model, the reverse shock heats this material to temperatures above $5\times10^9$~K, which is the upper limit for incomplete silicon burning. Thus, incomplete silicon burning abruptly ceases inside the oxygen burning shell, as it is replaced by complete silicon burning. This shift to complete silicon burning can be seen in Figure \ref{fig:nucleozooms} (\textsl{middle}), where we see that chromium production is replaced by iron production for explosion energy above $\sim$ 5.75 B. \\

The reason that the reverse shock has no significant effect on nucleosynthesis in the $40\,\Msun$, $60\,\Msun$, and $80\,\Msun$ models is due to a combination of the larger oxygen/neon burning shell in the initial model, and the less significant change in density between shells for these masses in comparison to the $15\,\Msun$, $20\,\Msun$, and $30\,\Msun$ models. This is best visualised in Figures \ref{fig:z15_shockconditions} to \ref{fig:z80_shockconditions}, which show the $\rho r^3$ profile of each initial model.

\onecolumn

\twocolumn

In each of these figures, the piston location is shown by the vertical dashed line, and the discontinuities in $\rho r^3$ reveal the shell interfaces. If silicon burning reaches a shell interface, and the reverse shock launched from the density discontinuity is large enough, then we find the nucleosynthesis process to be significantly affected, as we have described for the $15\,\Msun$ model. The $15\,\Msun$ model is representative of both the $20\,\Msun$ and $30\,\Msun$ models, and we find a significant reverse shock to be launched from the shell interface at $m\simeq1.65\,\Msun$, $m\simeq1.85\,\Msun$, and $m\simeq2.25\,\Msun$, for each of these models respectively. This is due to the close proximity of these shell interfaces to the piston, and the sharpness of each discontinuity, i.e., how abruptly the shock decelerates. In the $40\,\Msun$, $60\,\Msun$, and $80\,\Msun$ models, we see that there is increasingly more mass between the piston and the nearest shell interface, and the discontinuity in $\rho r^3$ at the shell interface becomes smoother. We find this to result in a lower fraction of energy being imparted into reverse shock, and therefore a negligible effect on the nucleosynthesis process.

\subsection{Minimum Explosion Energy and Fallback}
\begin{figure}
	\includegraphics[width=\columnwidth]{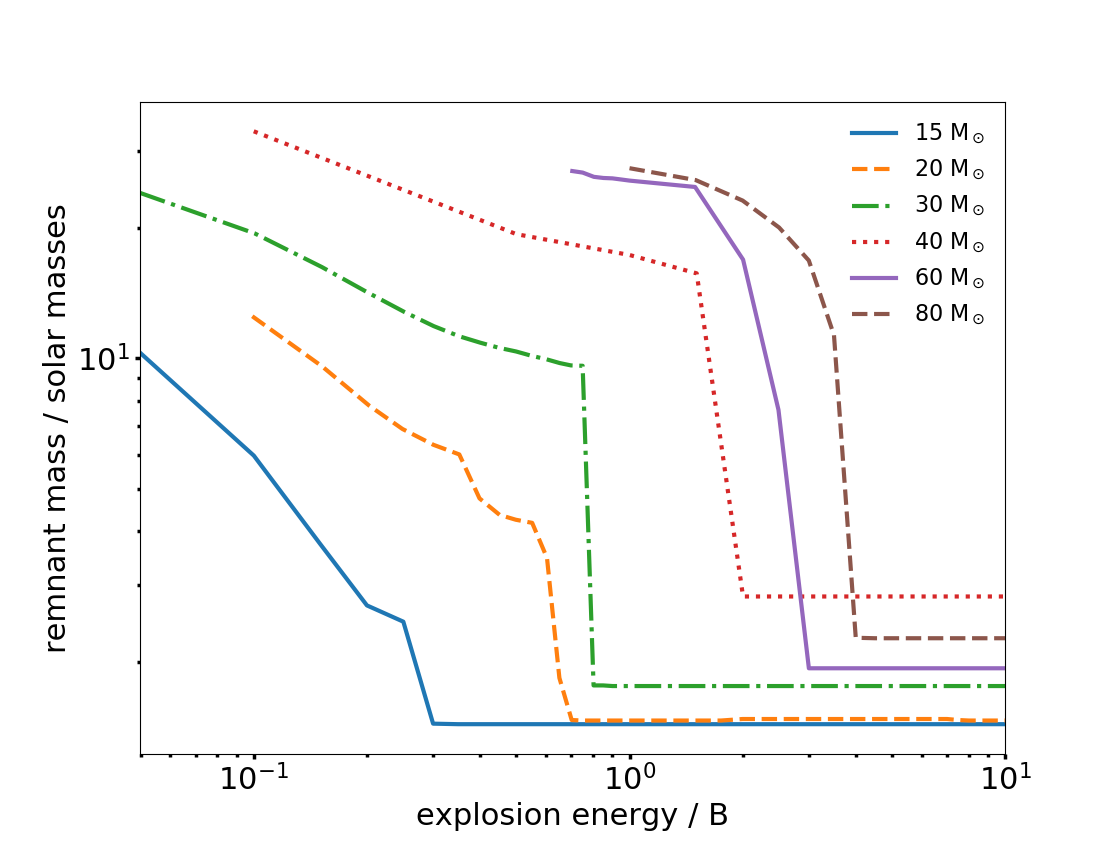}
    \caption{Central remnant mass as a function of explosion energy for each progenitor model.}
    \label{fig:remnant_mass}
\end{figure}

The explosion energy we calculate for each model is the final kinetic energy at infinity, and whenever we state "explosion energy", it is this energy that we are referring to. As such, the explosion mechanism itself must provide this final kinetic energy \textit{in addition} to the energy required to overcome the gravitational binding energy of the ejected material, as is described by \citet{woosley_1995}. If the energy supplied by the explosion mechanism is insufficient, some portion of the mantle will fall back onto the central remnant. In each model, we find a minimum threshold explosion energy that is required to eject all material above the piston, without any falling back. These minimum explosion energies are 0.3~B, 0.7~B, 0.8~B, 2~B, 3~B, and 4~B, for the $15\,\Msun$, $20\,\Msun$, $30\,\Msun$, $40\,\Msun$, $60\,\Msun$, and $80\,\Msun$ models, respectively. This can be seen in Figure \ref{fig:remnant_mass}. For the remainder of our results, we refer to these threshold explosion energies whenever we state the "minimum explosion energy." These results for remnant mass are in some disagreement with similar studies \citep[e.g.,][]{woosley_1995,zhang_2008}. It is likely that at low explosion energy, we are underestimating the amount of fallback. This is due to our piston acting as an artificial boundary condition, on top of which in-falling matter may pile up, as was also found by \citet{macfadyen_2001}. This is only an issue for lower explosion energy models, and for the high explosion energy models that we are interested in, it will not play a role. The remnant masses we achieve for higher explosion energy are consistent with the results of similar studies, as we discuss in Section \ref{sec:discussion}. Nevertheless, this effect should be kept in mind when considering our results for the lowest explosion energies without fallback in each model.

\subsection{Iron-Peak Elements}
The final abundances of the iron-peak elements as a function of energy are shown in Figures \ref{fig:z15_fepeak} - \ref{fig:z80_fepeak}. 

\subsubsection{Chromium}\label{sssec:cr}
The dominant channel of chromium production is incomplete silicon burning. By this process, chromium is synthesised primarily as $^{52}$Fe, and in small parts $^{53}$Fe, decaying to $^{52}$Cr and $^{53}$Cr, respectively.

Referring to Figures \ref{fig:z15_fepeak} to \ref{fig:z30_fepeak}, we see the relationship between [Cr/Fe] and explosion energy is qualitatively similar for the $15\,\Msun$, $20\,\Msun$, and $30\,\Msun$ models. We find that [Cr/Fe] remains approximately constant, at a value close to zero, for explosion energies less than 7~B, 5.5~B, and 9.5~B, in each of these models respectively. For explosion energies less than these, chromium production occurs well within the oxygen/neon burning shell, as we can see in Figures \ref{fig:z15_totalnucleo} to \ref{fig:z30_totalnucleo}. At an explosion energy of approximately 7~B, 5.5~B, and 9.5~B in each of these models respectively, chromium production nears the outer edge of the oxygen/neon burning shell (labelled in Figures \ref{fig:z15_totalnucleo} to \ref{fig:z30_totalnucleo}), where the reverse shock described in Section \ref{ssec:rev_shock} results in a dip in the [Cr/Fe] value, reaching a minimum of approximately $-0.2$, $-0.3$, and $-0.2$, for the $15\,\Msun$, $20\,\Msun$, and $30\,\Msun$ models, respectively. For explosion energies greater than this, [Cr/Fe] increases, and reaches a maximum of $\sim0.15$ in each of these models for our greatest explosion energy, 100 B.\\  
For the $40\,\Msun$ model (Figure \ref{fig:z41_fepeak}), the change in [Cr/Fe] with explosion energy looks very similar to that seen in lower mass models, remaining approximately constant at -0.1 until an explosion energy $\sim 8 \,\mathrm{B}$, where [Cr/Fe] begins to increase, again reaching a maximum of $\sim 0.15$ for explosion energy 100~B. In this case, and for all models greater than $30\,\Msun$, the reverse shock has no noticeable effect on the final chemical yields of silicon burning, i.e., there is no dip in [Cr/Fe] when silicon burning occurs at the outer edge of the oxygen/neon burning shell.
For the $60\,\Msun$ and $80\,\Msun$ models (Figures \ref{fig:z60_fepeak} and \ref{fig:z80_fepeak}), [Cr/Fe] shows little change for explosion energies up to $\sim 10 \,\mathrm{B}$, remaining close to [Cr/Fe]~$\simeq 0.0$, and \textit{decreasing} to values slightly less than zero for greater explosion energies.\\
At high explosion energy ($\gtrsim 10 \, \mathrm{B}$), we have found an increase in [Cr/Fe] for the $15\,\Msun$ -- $40\,\Msun$ models, and a decrease in [Cr/Fe] in the $60\,\Msun$ and $80\,\Msun$ models. The reason behind this is not as much to do with chromium synthesis, but more so with iron production. For increasing explosion energy, we find that the innermost regions of the star begin to be dominated by $\alpha$-particle production, in regions that would otherwise produce iron, as can be seen in Figures \ref{fig:z15_totalnucleo} to \ref{fig:z80_totalnucleo}. This is due to an increased expansion time of the material, as is described in Section \ref{ssec:alpharich}. For a given post-shock temperature, the extent to which $\alpha$-particle abundance dominates is dependent on the density of the material. The lower mass models ($15\,\Msun$ to $40\,\Msun$) have lower densities, which allows for a stronger $\alpha$-rich freezeout for any given explosion energy, which reduces iron ($^{56}$Ni) production, and leads to an increase in [Cr/Fe]. The $60\,\Msun$ and $80\,\Msun$ models have greater densities, which reduces the effect of an $\alpha$-rich freezeout. In these models iron (as $^{56}$Ni) is still produced throughout the majority of the inner shells for high explosion energy, and [Cr/Fe] decreases in turn.\\
We are not able to reproduce the low values of [Cr/Fe] that are observed in metal poor stars, i.e., $\sim-0.5$ \citep{cayrel_2004}. We achieve the largest deviation from [Cr/Fe]$\simeq 0$ in models for which the reverse shock at the inner shell interface most affects incomplete silicon burning. In these models, we find [Cr/Fe] values between $-$0.2 -- $-$0.3

\subsubsection{Manganese}\label{sssec:mn}
Manganese is synthesised during incomplete silicon burning, primarily as $^{55}$Co, and in small parts $^{55}$Fe. Both decay to $^{55}$Mn.
For the $15\,\Msun$, $20\,\Msun$, and $30\,\Msun$ models, we find that [Mn/Fe] decreases steadily with increasing explosion energy, up until explosion energies of approximately 7 B, 5.5 B, and 9.5 B, respectively, as can be seen in Figures \ref{fig:z15_fepeak} to \ref{fig:z30_fepeak}. The initial decrease in [Mn/Fe] as silicon burning moves outward for increasing explosion energy is due to a decreased neutron excess at larger mass coordinate, as is described in Section \ref{ssec:neutron_excess}. $^{55}$Co requires an excess of neutrons to be synthesised, so as silicon burning moves outwardly for increasing explosion energy, $^{55}$Co synthesis is constrained by the neutron excess, and the final manganese yield is decreased. At explosion energies 7 B, 5.5 B, and 9.5 B, for the $15\,\Msun$, $20\,\Msun$, and $30\,\Msun$ models respectively, manganese production occurs at the outer edge of the oxygen/neon burning shell, and the reverse shock reduces manganese synthesis, as can be seen in Figures \ref{fig:z15_totalnucleo} to \ref{fig:z30_totalnucleo}, decreasing [Mn/Fe] to a value of $-0.8$, $-0.85$, and $-1.0$ for the $15\,\Msun$, $20\,\Msun$, and $30\,\Msun$ models, respectively. 
In Figures \ref{fig:z15_fepeak} to \ref{fig:z30_fepeak}, it can be seen that for explosion energies immediately larger than this, unlike [Cr/Fe], [Mn/Fe] does not immediately recover to the value it took for lower explosion energies, where manganese synthesis occurred well inside the oxygen/neon burning shell. This is a result of an abrupt decrease in the neutron excess between shells, as can be seen in Figures \ref{fig:z15_shockconditions} to \ref{fig:z80_shockconditions}. For explosion energies greater than 7 B, 5.5 B, and 9.5 B in the $15\,\Msun$, $20\,\Msun$, and $30\,\Msun$ models, respectively, [Mn/Fe] shows an increase, which is again due to a decrease in iron production as a result of a strong $\alpha$-rich freezeout at high explosion energy in these models.\\
We find a consistent decrease in [Mn/Fe] for increasing explosion energy in the $40\,\Msun$, $60\,\Msun$, and $80\,\Msun$ models (Figures \ref{fig:z41_fepeak} to \ref{fig:z80_fepeak}). This is due to a combination of the outwardly decreasing density and neutron excess, which limits manganese synthesis, and only a moderate $\alpha$-rich freeze-out in these models at high explosion energy, which allows for increasing iron production.\\
We are able to achieve the low values of [Mn/Fe] which are observed in metal poor stars, being [Mn/Fe]$\simeq-0.5$, with several models \citep{cayrel_2004}. We achieve $-0.25 \geq \mathrm{[Mn/Fe]} \geq -1$, for all models, with the exception of the $60\,\Msun$ and $80\,\Msun$ models for explosion energy greater than 30~B -- 40~B, where [Mn/Fe] falls below $-1$. The models with [Mn/Fe] closest to $-0.5$ are the $15\,\Msun$, $20\,\Msun$, and $30\,\Msun$, with explosion energies $\sim$5~B, and the $60\,\Msun$ and $80\,\Msun$ models, with explosion energies $\lesssim 15$~B.

\subsubsection{Cobalt}
The dominant channel of cobalt production is complete silicon burning, where cobalt is primarily synthesised as $^{59}$Ni and $^{59}$Cu, which both decay to $^{59}$Co.
We find that an increasing explosion energy typically serves to increase [Co/Fe], with the exception of the cases for which either cobalt synthesis does not occur in any significant amount for low explosion energy, or there is an extremely strong $\alpha$-rich freezeout in the inner regions for high explosion energy.
For the $15\,\Msun$ model, we find that cobalt synthesis occurs in the ejected layers even for our lowest explosion energies, so that [Co/Fe] increases immediately with explosion energy above our minimum energy, as can be seen in Figure \ref{fig:z15_fepeak}. While we do find that $\alpha$-rich freezeout dominates the innermost regions of the star for high explosion energy, cobalt also is increasingly synthesised in a second region, within the carbon burning shell, for explosion energy $\gtrsim 30$~B (Figure \ref{fig:z15_totalnucleo}). This is sufficient to compensate for the reduced cobalt synthesis in the oxygen/neon burning shell, allowing for a continuing increase of [Co/Fe] with explosion energy. In the $20\,\Msun$ model, we find that [Co/Fe] remains approximately constant for explosion energies less than 4~B, as seen in Figure \ref{fig:z20_fepeak}. This is due to only a small amount of cobalt being synthesised in the ejected layers for our lowest explosion energies, as can be seen in Figure \ref{fig:z20_totalnucleo}. For explosion energy greater than 3~B -- 4~B, synthesis of cobalt increases and shifts outward, and [Co/Fe] increases with explosion energy. X(He) begins to exceed 0.5 in the inner regions for explosion energy greater than $\sim$~10~B, which restricts the cobalt yield, and [Co/Fe] reaches a plateau for our greatest explosion energies. For the $30\,\Msun$ model (Figure \ref{fig:z30_fepeak}), cobalt synthesis occurs for our lowest explosion energies, and increases with explosion energy. Similar to the $20\,\Msun$ model, the effect of an extremely strong $\alpha$-rich freezeout can be seen as the [Co/Fe] value reaches a plateau for our highest explosion energies. In the $40\,\Msun$ model, we find a steady increase of [Co/Fe] with explosion energy (Figure \ref{fig:z41_fepeak}), as cobalt production occurs for all explosion energies, and the higher densities in this model result in only a moderate $\alpha$-rich freezeout. In the $60\,\Msun$ and $80\,\Msun$ models, we find only relatively small amounts of cobalt production in the ejected layers for explosion energies less than $\sim30$~B and $\sim35$~B, respectively, as can be seen in Figures \ref{fig:z60_totalnucleo} and \ref{fig:z80_totalnucleo}. Combined with increasing iron synthesis throughout the inner layers, we find a decreasing value of [Co/Fe] for explosion energies less than these values, as can be seen in Figures \ref{fig:z60_fepeak} and \ref{fig:z80_fepeak}. For greater explosion energies, cobalt is synthesised in increasingly larger amounts in the inner layers as the silicon burning regions are pushed outwards, and [Co/Fe] increases for greater explosion energy.\\
We are able to achieve the abundance typically observed in metal-poor stars, i.e., [Co/Fe]~$\simeq 0.5$ \citep{cayrel_2004} in the $15\,\Msun$ for explosion energy 20~B -- 30~B, and with the $20\,\Msun$, and $30\,\Msun$ models with explosion energy greater than $\sim$20~B.

\subsubsection{Zinc}
The dominant channel of zinc synthesis is complete silicon burning. Zinc is primarily synthesised as $^{64}$Ga and $^{64}$Ge, which both decay to $^{64}$Zn, and also in smaller parts directly as $^{64}$Zn. $^{66}$Ge, decaying to $^{66}$Zn, is also produced in small amounts throughout the complete silicon burning region. 
An increase in explosion energy broadens the region producing zinc, and increases [Zn/Fe]. Similar to the way we described for cobalt, the only cases where [Zn/Fe] does not increase with explosion energy is for low ($\lesssim$~4~B) explosion energy in the $20\,\Msun$ model, where zinc production borders the mass cut, and high ($\gtrsim$~30~B) explosion energy in the $15\,\Msun$, $20\,\Msun$, and $30\,\Msun$ models, where the final abundance of the innermost ejecta is dominated by helium as a consequence of the fast expansion time of the shocked material, as can be seen in Figures \ref{fig:z15_totalnucleo} to \ref{fig:z30_totalnucleo}. 

\onecolumn
\begin{figure}
    \centering
    \begin{subfigure}[b]{0.5\textwidth}
        \centering\captionsetup{width=.8\linewidth}
        \includegraphics[width=\textwidth]{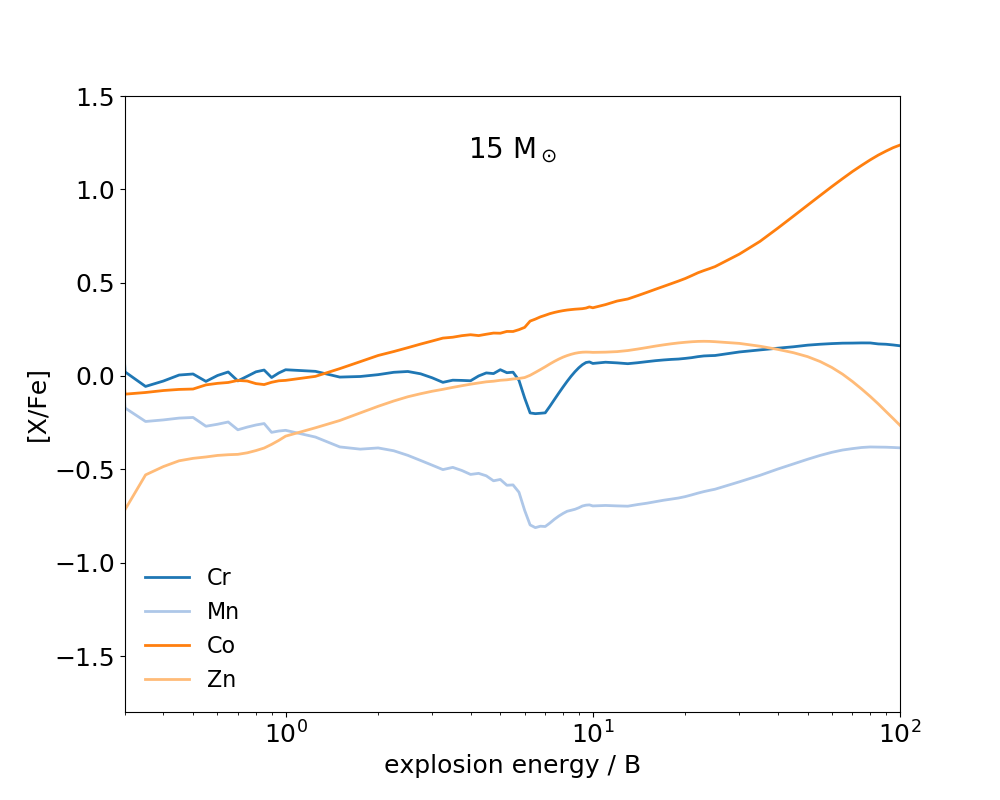}
		\caption{The $15\,\Msun$ model.}
		\label{fig:z15_fepeak}
    \end{subfigure}%
    \begin{subfigure}[b]{0.5\textwidth}
        \centering\captionsetup{width=.8\linewidth}
        \includegraphics[width=\textwidth]{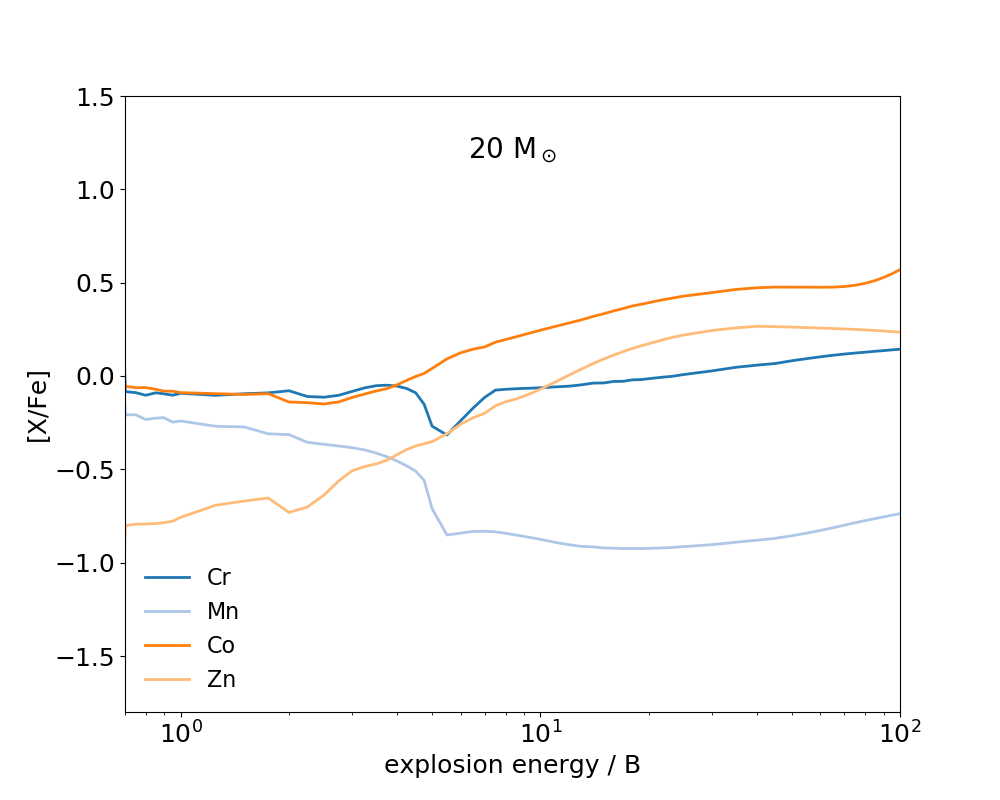}
		\caption{The $20\,\Msun$ model.}
		\label{fig:z20_fepeak}
    \end{subfigure}
    \begin{subfigure}[b]{0.5\textwidth}
        \centering\captionsetup{width=.8\linewidth}
        \includegraphics[width=\textwidth]{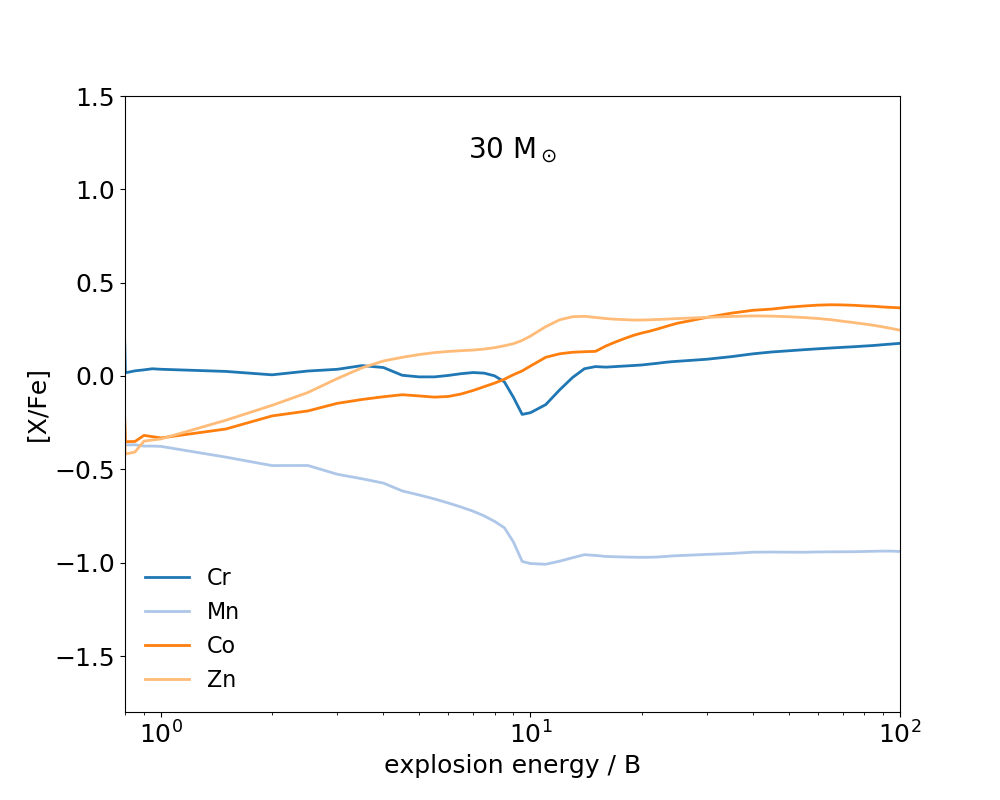}
		\caption{The $30\,\Msun$ model.}
		\label{fig:z30_fepeak}
    \end{subfigure}%
    ~ 
    \begin{subfigure}[b]{0.5\textwidth}
        \centering\captionsetup{width=.8\linewidth}
        \includegraphics[width=\textwidth]{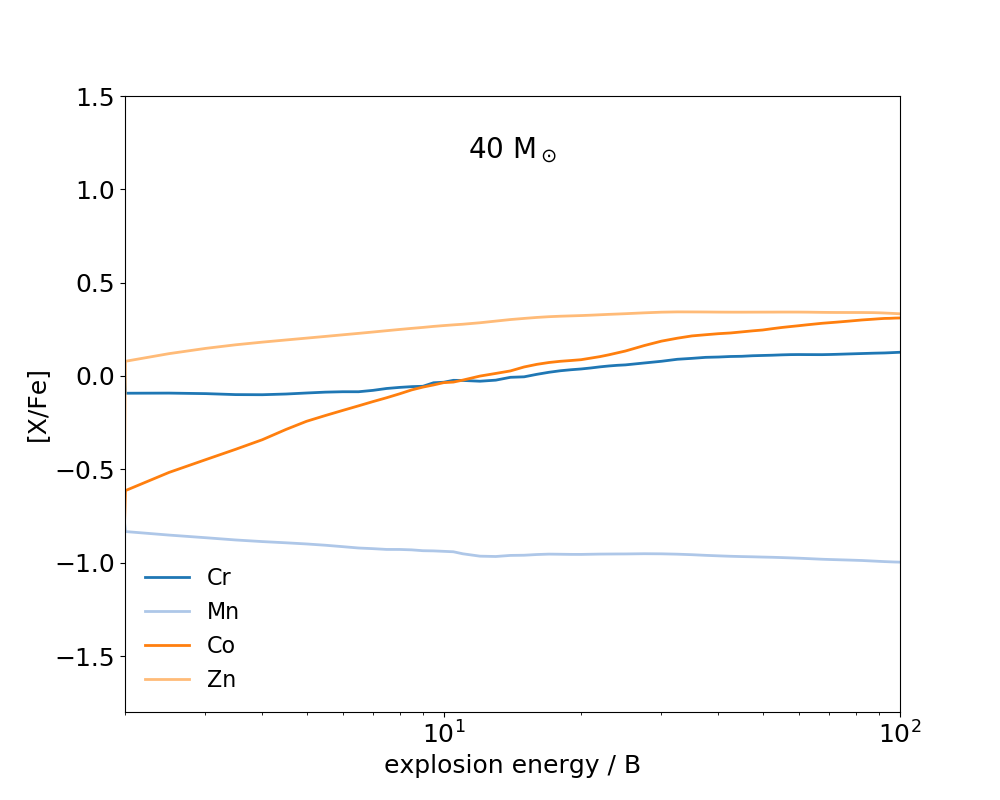}
		\caption{The $40\,\Msun$ model.}
		\label{fig:z41_fepeak}
    \end{subfigure}
     \begin{subfigure}[b]{0.5\textwidth}
        \centering\captionsetup{width=.8\linewidth}
        \includegraphics[width=\textwidth]{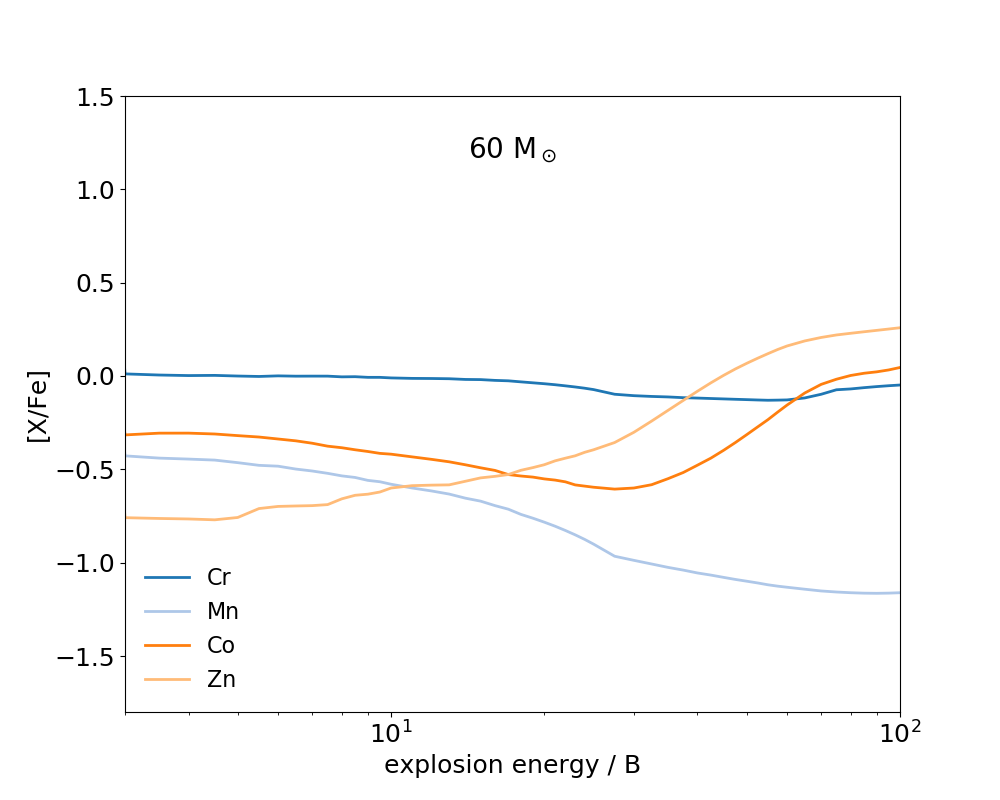}
		\caption{The $60\,\Msun$ model.}
		\label{fig:z60_fepeak}
    \end{subfigure}%
    ~ 
    \begin{subfigure}[b]{0.5\textwidth}
        \centering\captionsetup{width=.8\linewidth}
        \includegraphics[width=\textwidth]{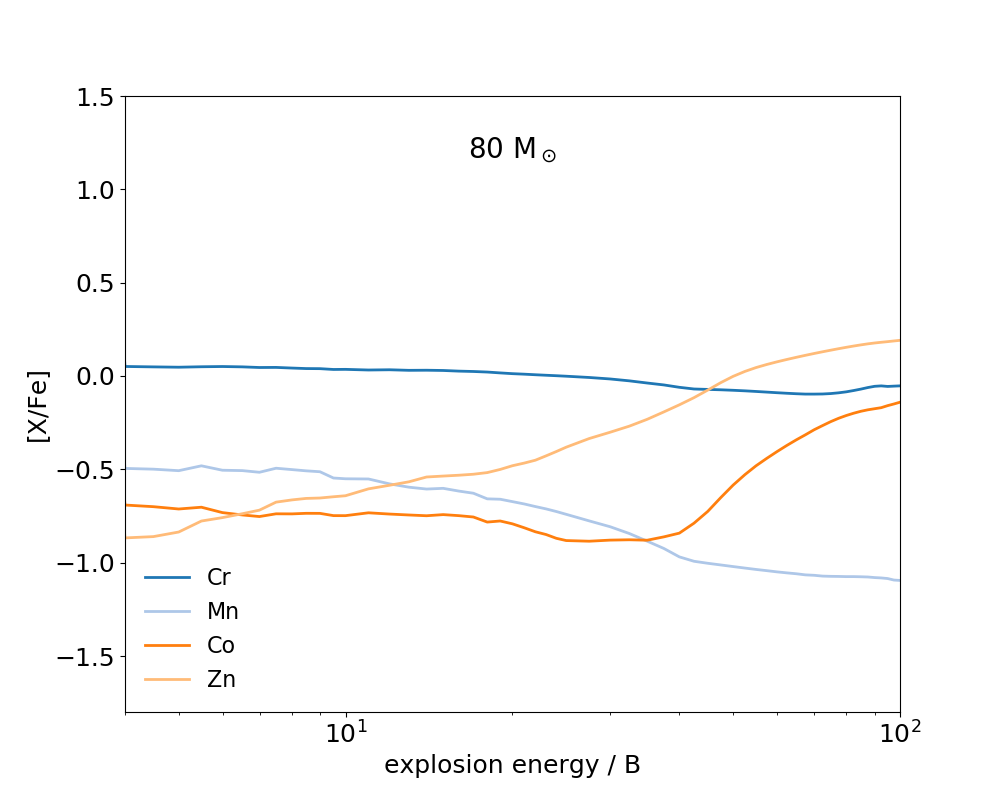}
		\caption{The $80\,\Msun$ model.}
		\label{fig:z80_fepeak}
    \end{subfigure}
    \caption{Abundances of iron peak elements in ejecta relative to the solar value, for each of the models $15\,\Msun$ -- $80\,\Msun$.}
\end{figure}

\twocolumn

While a moderate $\alpha$-rich freezeout can enhance zinc production, because $^{64}$Ge is synthesised through the capture of $\alpha$-particles, a continued increase in explosion energy can result in the material expanding too fast for $\alpha$-particles to recombine at all, and we find local X(He)~>~0.60. This is detrimental to zinc ($^{64}$Ge) synthesis, as can be seen in Figure \ref{fig:z15_totalnucleo}. The degree of $\alpha$-rich freezeout becomes less significant as progenitor mass increases due to the higher density, and therefore increasing expansion time in more massive stars. In the $20\,\Msun$ and $30\,\Msun$ models the [Zn/Fe] value plateaus for explosion energy greater than 20~B -- 30~B, but does not show any significant decrease for higher explosion energy, as it does for the $15\,\Msun$ model (Figures \ref{fig:z15_fepeak} to \ref{fig:z30_fepeak}). For the $40\,\Msun$ model we find the least variation in [Zn/Fe] for all models, shown in Figure \ref{fig:z41_fepeak}, as zinc is produced all explosion energies that eject the inner regions, and the degree of $\alpha$-rich freezeout remains moderate for our largest explosion energies, which is beneficial to zinc production, as can be seen in Figure \ref{fig:z41_totalnucleo}. For the $60\,\Msun$ and $80\,\Msun$ models, we find significant increase in [Zn/Fe] with explosion energy (Figures \ref{fig:z60_fepeak} and \ref{fig:z80_fepeak}). In these models, there are multiple sites of zinc synthesis in the inner layers at explosion energy $\gtrsim 30$~B, contribute to a large zinc yield, as can be seen in Figures \ref{fig:z60_totalnucleo} and \ref{fig:z80_totalnucleo}. \\
We are unable to reproduce the very high [Zn/Fe]~$\gtrsim 0.5$ observed in the most metal poor stars \citep{cayrel_2004,skuladottir_2017}. At most, we can achieve [Zn/Fe]~$\simeq 0.3$ in all models at high explosion energy. The exception is the $15\,\Msun$ model, where [Zn/Fe] barely reaches a maximum of 0.2 at an explosion energy of $\sim 25\,$B, due to $\alpha$-particles dominating the abundance of the inner regions at explosion energies larger than this.

\subsection{Intermediate-Mass Elements}

\subsubsection{Potassium}
We find that potassium is synthesised by two separate burning processes, and the ratio of the contributions from each to the total potassium yield is dependent on the progenitor mass, and the explosion energy. For the $15\,\Msun$, $20\,\Msun$, and $30\,\Msun$ models, potassium is predominantly synthesised as $^{39}$K via oxygen burning, for all explosion energies. $^{39}$K is neutron rich, and hence is produced in larger abundance when the fuel for burning contains excess neutrons. The neutron excess profile of the progenitor models decreases outwardly. Consequently, we see that [K/Fe] initially decreases with increasing explosion energy (Figures \ref{fig:z15_intmass} to \ref{fig:z30_intmass}) as explosive oxygen burning shifts into the outer shells, where the lower neutron excess inhibits the synthesis of $^{39}$K. In these models, oxygen burning commences at the very outer edge of the oxygen/neon burning shell for the lowest explosion energies. This shell has a large neutron excess in our models, hence [K/Fe] decreases significantly as oxygen burning shifts out of this shell.
For explosion energy of order $\sim10$~B in these models, [K/Fe] begins to increase for two reasons. The first is that a strong $\alpha$-rich freezeout begins to restrict iron production, increasing [K/Fe]. The second is that $^{41}$Ca begins to be produced in the innermost region of the ejecta. This region undergoes complete silicon burning and is left with the largest degree of $\alpha$-rich freezeout. In this case, the $^{41}$Ca is synthesised in regions with X(He)$\gtrsim0.85$. The $^{41}$Ca then decays to $^{41}$K via electon capture. The $^{41}$K contribution to the total potassium yield is small in these models, i.e., approximately an order of magnitude less than the $^{39}$K yield, though it does continue to increase for larger explosion energy.\\
For the $40\,\Msun$ model, we find a constant increase in [K/Fe] with explosion energy, as seen in Figure \ref{fig:z41_intmass}. In this model, larger explosion energies are required to eject the inner regions, which places oxygen burning at a more exterior mass coordinate, in a region with already relatively low neutron excess. Hence, [K/Fe] is already low for our lowest explosion energies. [K/Fe] then increases as faster expansion times limits iron production, and $^{41}$K begins to be produced in the innermost region as we described for the lower mass models. \\
For the $60\,\Msun$ and $80\,\Msun$ models, [K/Fe] decreases with increasing explosion energy up to 30~B -- 40~B, as oxygen burning moves outward into regions of decreasing neutron excess, limiting $^{39}$K synthesis (Figures \ref{fig:z60_intmass} and \ref{fig:z80_intmass}). Also, iron is produced in large amounts as the $\alpha$-rich freezeout remains moderate in these higher density models, further contributing to a low [K/Fe] value. For greater explosion energy, [K/Fe] begins to increase, as $^{41}$K production becomes significant at very high explosion energy. In these higher mass models, the $^{39}$K and $^{41}$K contributions to the total potassium abundance begin to converge. Though $^{39}$K remains slightly more abundant, even at an explosion energy of 100~B.\\
The only explosions in which we achieve [K/Fe] close to the value observed in metal-poor stars (i.e., [K/Fe]$\simeq0$, \citet{cayrel_2004}), is with explosions of energy $<1$~B in the $15\,\Msun$ model, where K is synthesised in abundance in the neutron rich oxygen-neon burning shell.
 
\subsubsection{Scandium}\label{sssec:sc}
We first must note that the sporadic "spikes" that can be seen in [Sc/Fe], particularly for explosion energies below 10~B in the $30\,\Msun$ and $80\,\Msun$ models (Figures \ref{fig:z30_intmass} and \ref{fig:z80_intmass}), are only due to the limited mass resolution of our calculations. Scandium is produced in a very narrow region, and we do not have adequate mass resolution to reliably track the synthesis of scandium in these models. For all other regimes our mass resolution seems to be sufficient, but the [Sc/Fe] in the $30\,\Msun$ and $80\,\Msun$ models for explosion energy less than 10~B should not be taken as reliable results. \\
We find that scandium is produced predominately in the complete silicon burning region as $^{45}$Ti, which is enhanced by a moderate $\alpha$-rich freezeout, and decays to $^{45}$Sc. With the exception of a few cases to be described, we find that [Sc/Fe] shows a strong positive correlation with explosion energy. The first regime of decreasing [Sc/Fe] for increasing explosion energy occurs for the lowest explosion energies in all models other than the $40\,\Msun$ model. We find that the majority of scandium synthesis falls inside the mass cut for our lowest explosion energies. Increasing explosion energy serves to increase iron production. This results in low [Sc/Fe] values for these models, until such an explosion energy for which scandium production shifts outside of the mass cut. This is clearly seen in Figures \ref{fig:z15_intmass} to \ref{fig:z80_intmass}. In the $40\,\Msun$ model, scandium production falls outside of the mass cut for our lowest explosion energies, and this effect is not seen (Figure \ref{fig:z41_intmass}). The other regime in which increasing explosion energy results in a decrease in [Sc/Fe] is for the $15\,\Msun$ model (and to a very minor extent the $20\,\Msun$ and $30\,\Msun$ models), where high ($\gtrsim 30$~B) explosion energy results in $\alpha$-particle production dominating the inner regions where scandium would otherwise be produced.\\
We are able to achieve [Sc/Fe]~$\simeq$~0, which is the observed  in metal-poor stars \citep{cayrel_2004}, with explosions of order 10~B, in all models with progenitor mass $\leq30\,\Msun$. We can achieve similar results in the $40\,\Msun$ model for higher explosion energies, closer to 100~B, but find that [Sc/Fe] barely reaches $-0.5$ in the $60\,\Msun$ and $80\,\Msun$ models, even for an explosion energy of 100~B. The low [Sc/Fe] in these models is predominately due to the high iron production in these higher density models.

\subsubsection{Titanium}
Titanium is synthesised throughout both the complete and incomplete silicon burning regions, primarily as $^{48}$Cr, and in smaller amounts as $^{49}$Cr. These isotopes decay to $^{48}$Ti and $^{49}$Ti, respectively.
Referring to Figures \ref{fig:z15_intmass} to \ref{fig:z80_intmass} we see that, generally, [Ti/Fe] shows a weak positive correlation with explosion energy for all progenitor masses. There are two regimes where this does not hold. The first is for explosion energies greater than $\sim$10~B in the $15\,\Msun$ model, where $\alpha$-particle production begins to dominate the inner regions to an extent which limits titanium synthesis (this affects iron production too of course, but titanium production is reduced more significantly). The second is for explosion energy less than $\sim$10~B in the $80\,\Msun$ model. Here, titanium is produced more abundantly in the innermost region, where the neutron excess is high. [Ti/Fe] then decreases slightly with explosion energy up to $\sim$10~B, as titanium synthesis shifts outwards. [Ti/Fe] then remains approximately constant for higher explosion energy.\\ 
For the $15\,\Msun$, $20\,\Msun$, and  $30\,\Msun$ models, for explosion energies which place silicon burning in the vicinity of the inner shell boundary, the effect of reverse shock heating can be seen in the [Ti/Fe] value, as has been discussed for the iron-peak elements. The effect of reverse shock heating for [Ti/Fe] is slightly less than for [(Cr,Mn)/Fe]. This is  because titanium is produced broadly throughout the entire silicon burning region, and the small region which is affected by the reverse shock at the oxygen burning/carbon burning shell interface has little impact on the total titanium yield. \\
The large [Ti/Fe] observed in metal-poor stars, approaching 0.5, cannot be reproduced with these models \citep{cayrel_2004}. While we can achieve [Ti/Fe] approaching 0.25 in models $\leq30\,\Msun$ for explosion energy $>10$~B, [Ti/Fe] remains below 0 for our range of explosion energies in models with progenitor mass $\geq40\,\Msun$. As for scandium, the low [Ti/Fe] in the higher mass models is predominantly due to a moderate $\alpha$-rich freezeout, which allows for high iron production.\\

\subsubsection{Vanadium}
Vanadium is produced predominately in the innermost complete silicon burning region, though also in smaller amounts in the incomplete silicon burning region. Vanadium is synthesised mainly as $^{51}$Mn, and in smaller amounts as $^{51}$Cr, which both decay to $^{51}$V. 

\begin{figure}
    \centering
    \begin{subfigure}[b]{0.5\textwidth}
        \centering\captionsetup{width=.8\linewidth}
        \includegraphics[width=\textwidth]{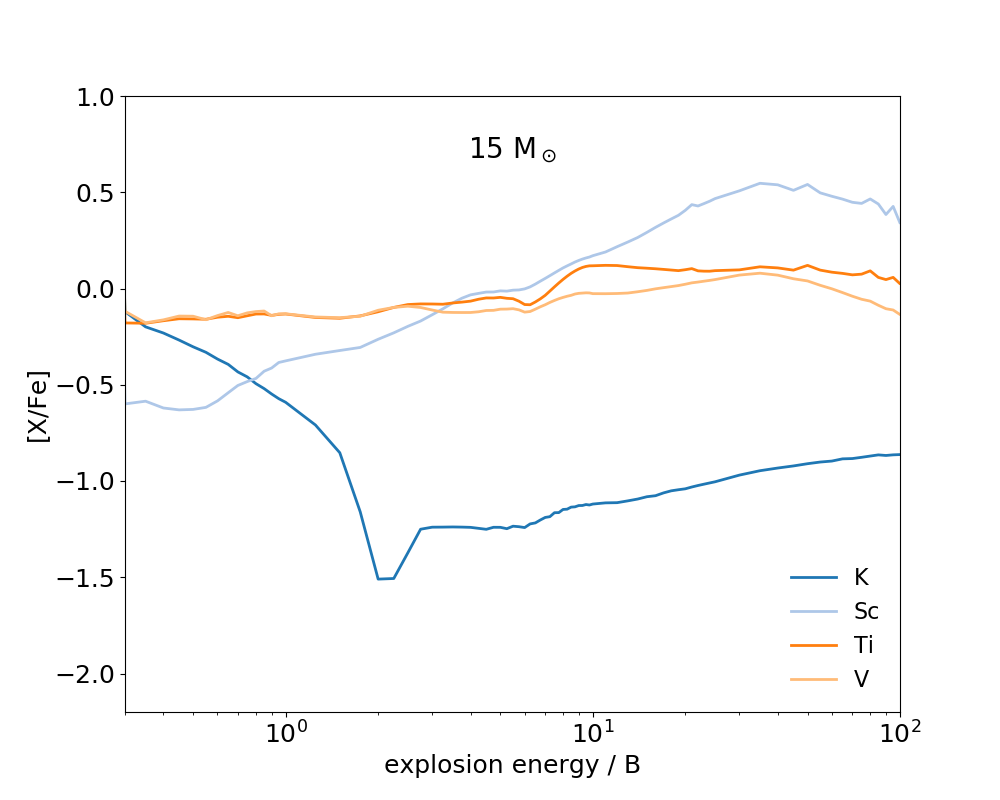}
		\caption{$15\,\Msun$}
		\label{fig:z15_intmass}
    \end{subfigure}%
    \begin{subfigure}[b]{0.5\textwidth}
        \centering\captionsetup{width=.8\linewidth}
        \includegraphics[width=\textwidth]{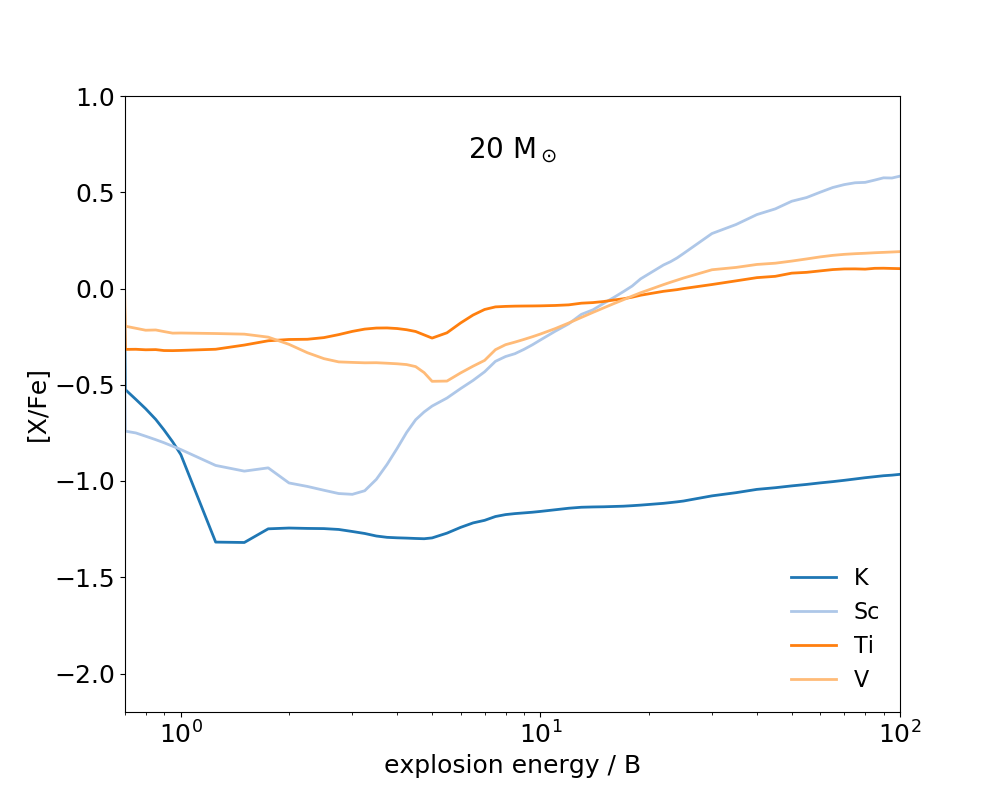}
		\caption{$20\,\Msun$}
		\label{fig:z20_intmass}
    \end{subfigure}
    \begin{subfigure}[b]{0.5\textwidth}
        \centering\captionsetup{width=.8\linewidth}
        \includegraphics[width=\textwidth]{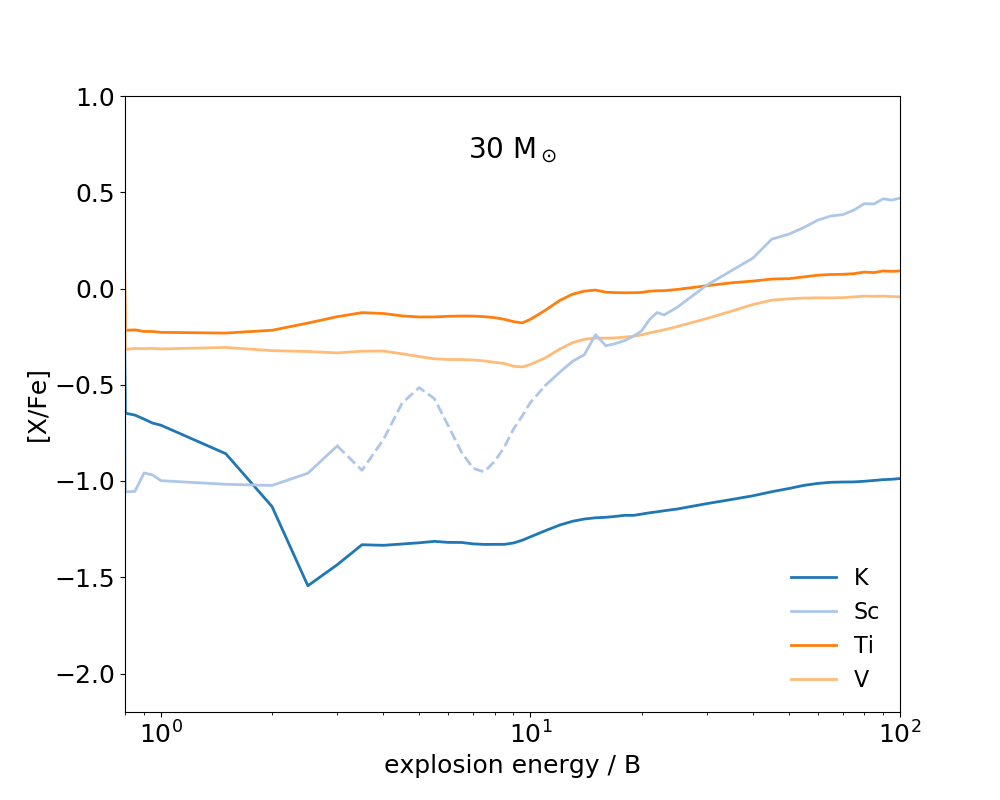}
		\caption{$30\,\Msun$}
		\label{fig:z30_intmass}
    \end{subfigure}%
    ~ 
    \begin{subfigure}[b]{0.5\textwidth}
        \centering\captionsetup{width=.8\linewidth}
        \includegraphics[width=\textwidth]{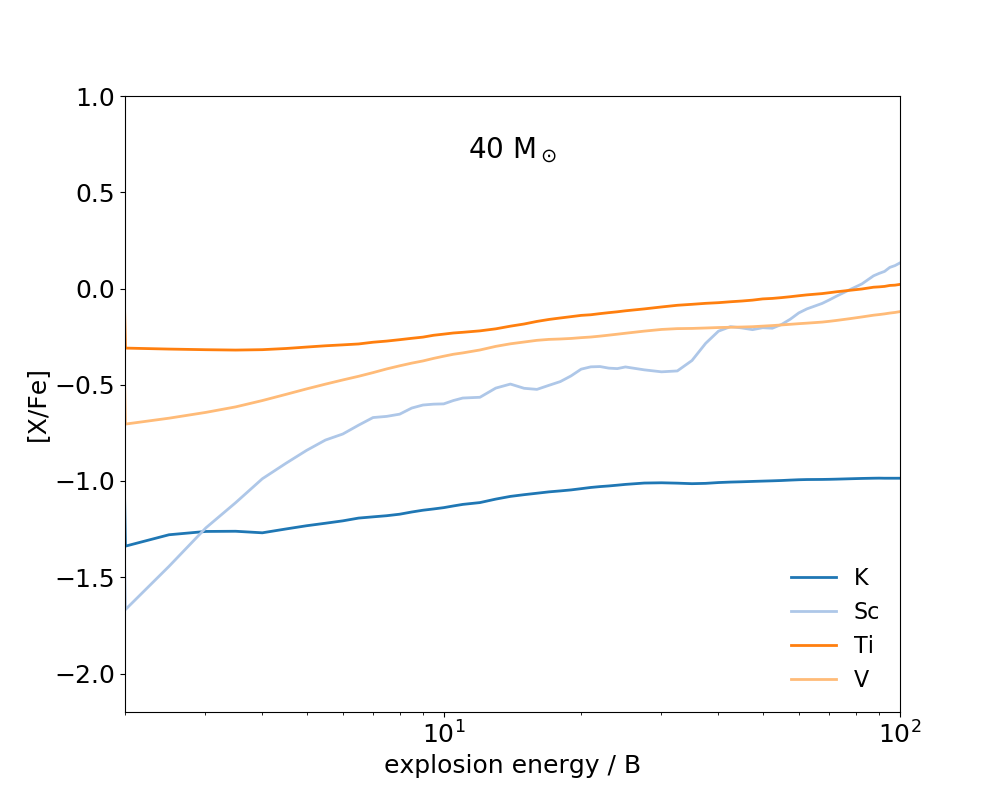}
		\caption{$40\,\Msun$}
		\label{fig:z41_intmass}
    \end{subfigure}
     \begin{subfigure}[b]{0.5\textwidth}
        \centering\captionsetup{width=.8\linewidth}
        \includegraphics[width=\textwidth]{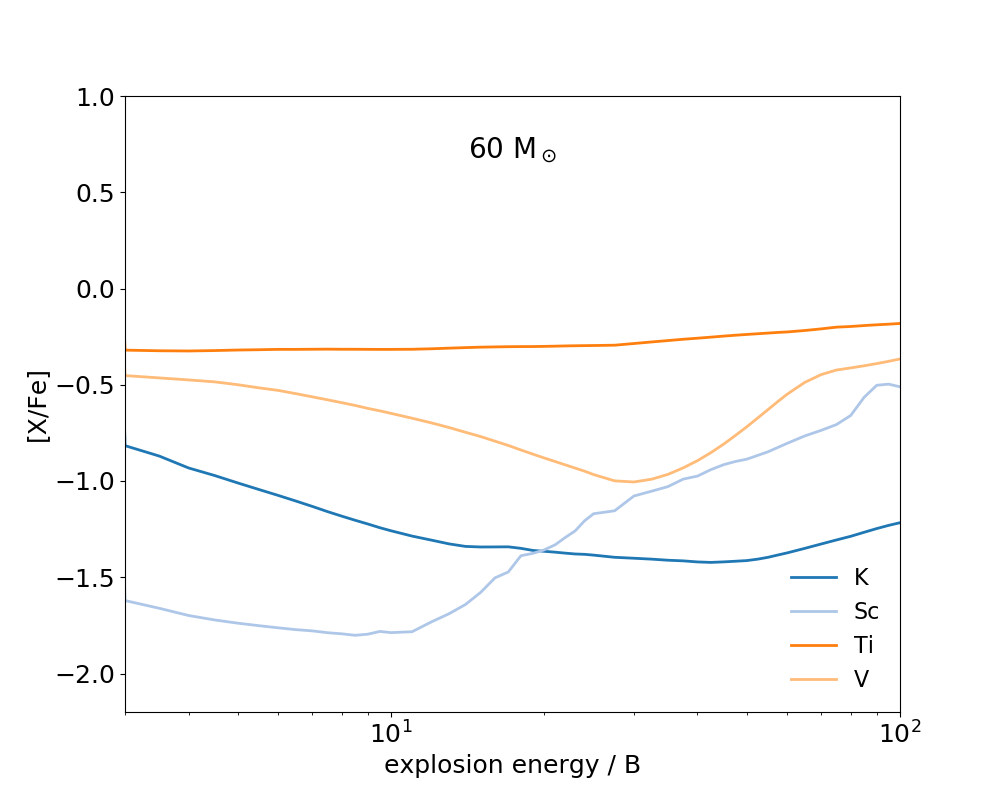}
		\caption{$60\,\Msun$}
		\label{fig:z60_intmass}
    \end{subfigure}%
    ~ 
    \begin{subfigure}[b]{0.5\textwidth}
        \centering\captionsetup{width=.8\linewidth}
        \includegraphics[width=\textwidth]{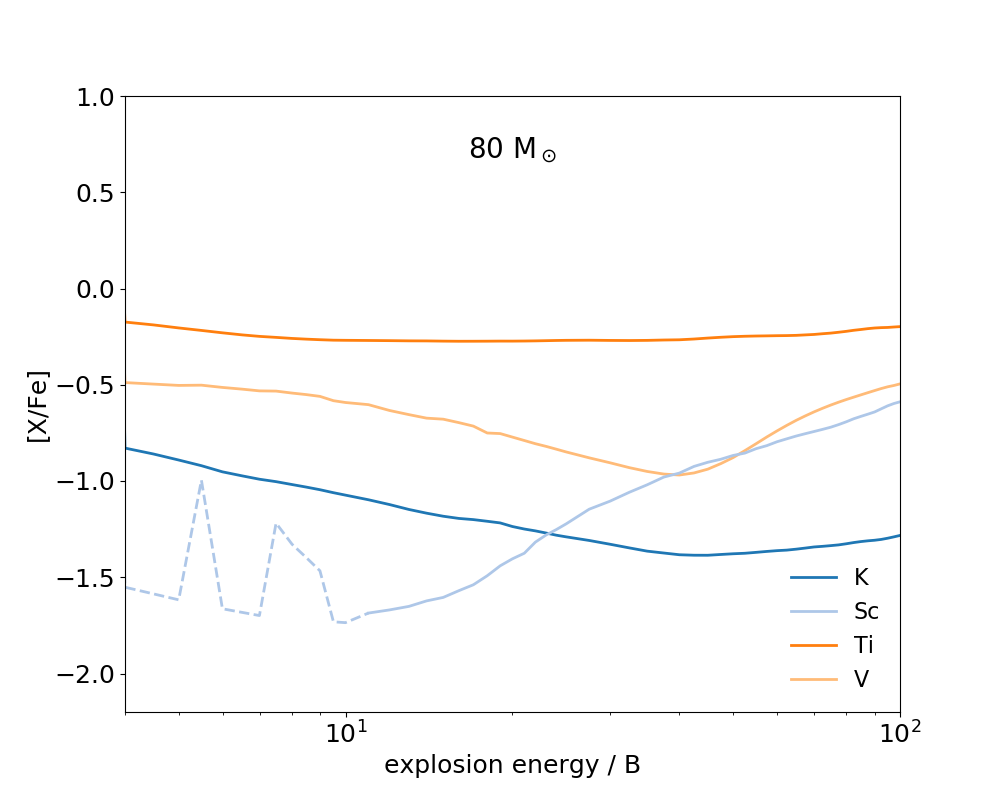}
		\caption{$80\,\Msun$}
		\label{fig:z80_intmass}
    \end{subfigure}
    \caption{Abundances of potassium, scandium, titanium, vanadium, in ejecta relative to the solar value, for each of the models $15\,\Msun$ -- $80\,\Msun$. The dashed line section for Sc in the $30\,\Msun$ and $80\,\Msun$ models indicate models for which we are not able to reliably track the synthesis of Sc due to limited mass resolution; see Section \ref{sssec:sc}.}
\end{figure}

For the $15\,\Msun$ model, vanadium synthesis straddles the mass cut for our lowest explosion energies, and an increasing explosion energy shifts a larger amount of vanadium synthesis into the ejected material. In Figure \ref{fig:z15_intmass}, we can see this effect as a steadily increasing [V/Fe] for increasing explosion energy. The exception to this trend is for explosion energies greater than 30~B -- 40~B, where fast expansion of the shocked inner regions results in X(He) increasing above $\sim0.7$, as can be seen in Figure \ref{fig:z15_totalnucleo}. This limits both vanadium and iron production in the innermost regions. Significant amounts of iron ($^{56}$Ni) continues to be produced at the outer edge of the complete silicon burning region even for our largest explosion energies, resulting in a decrease in [V/Fe] for explosion energies above 30~B -- 40~B in this model, as is seen in Figure \ref{fig:z15_intmass}.
For the $20\,\Msun$ model, most vanadium production lies completely inside of the mass cut for explosion energies $\lesssim5$~B, leaving only a small contribution to the vanadium yield in the incomplete silicon burning region. Iron ($^{56}$Ni) yield continues to increases with explosion energy, and the result of this is a decreasing [V/Fe] with increasing explosion energy for this range, as can be seen in Figure \ref{fig:z20_intmass}. Vanadium production only shifts outside of the mass cut in any significant amount for explosion energy at least $\sim 5$~B. For explosion energies greater than this [V/Fe] increases, as synthesis of vanadium extends through the ejected material. The [V/Fe] begins to plateau at explosion energies larger than 30~B -- 40~B, as can be seen for this energy range in Figure \ref{fig:z20_intmass}. This is again due to the increasingly fast expansion of the inner regions for greater explosion energy, which limits vanadium synthesis, as we saw for the $15\,\Msun$ model. Though, for more massive models and therefore larger internal densities (slower expansion times), this effect becomes less significant.
As can be seen in Figure \ref{fig:z30_intmass}, the relationship between [V/Fe] and explosion energy in the $30\,\Msun$ model is very similar to the $20\,\Msun$ model. Though, in the $30\,\Msun$ model, vanadium production begins to increase at an explosion energy of $\sim10$~B as opposed to $\sim5$~B in the $20\,\Msun$ model.

In the $40\,\Msun$ model, vanadium production begins well outside the mass cut for our minimum explosion energies, which allows a steady increase in [V/Fe] for increasing explosion energy, as can be seen in Figure \ref{fig:z41_intmass}. 

For the $60\,\Msun$ and $80\,\Msun$ models, no significant amounts of vanadium production occurs in the complete silicon burning region for explosion energies below $30\,$B -- $40\,$B, leaving only a small contribution to vanadium synthesis in the incomplete silicon burning region. Combined with large iron synthesis in these models, we find a small value of [V/Fe] in these models, which decreases with increasing explosion energy, until the explosion energy reaches $\sim$~40~B, as can be seen in Figures \ref{fig:z60_intmass} and \ref{fig:z80_intmass}. For higher explosion energies, vanadium begins to be synthesised in larger amounts within the complete silicon burning region, and the [V/Fe] value increases with explosion energy. \\

A large scatter of [V/Fe] values above $\sim$~0.2 are observed in metal poor stars \citep{honda_2004}. We can achieve such values in all models except for the $20\,\Msun$ model with explosion energies between $\sim$2~B -- 10~B, the $30\,\Msun$ model with explosion energies less than $\sim$10~B, the $40\,\Msun$ model with explosion energies less than $\sim$20~B, and in the $60\,\Msun$ and $80\,\Msun$ models for any explosion energy within our range.

\onecolumn

\twocolumn

\section{Aspherical Estimate}\label{sec:jets}
When calculating the results for a supernova using a one-dimensional code, it is necessary to make the assumption that the explosion is the same in all angular directions. The result calculated in one-dimension can then be integrated as a constant over the remaining two dimensions, making a three-dimensional result that is inherently spherical. Using our grid of one-dimensional models, which has a relatively fine explosion energy resolution, we can instead attempt a crude approximation that neglects lateral motions or mixing, and integrate over a weighted set of individual one-dimensional results to build up a non-spherical, multidimensional model. Hence we use the term "multidimensional" loosely, as each one-dimensional component of the final model is evolved independently of the next. In practice, we calculate this as follows; first, we assume some distribution of explosion energy with respect to the polar angle $\theta$. Here we will be taking a distribution with the energy focussed at the poles, representing jets launched in opposite directions. 

We define a simplisitic energy profile for our jets;

\[
    e(\theta) = \left\{\begin{array}{lr}
        e_j, & \text{for } \theta\leq \theta_c\\
        e_s, & \text{for } \theta> \theta_c
        \end{array}\right\},
  \]

where $e_j$ is the energy per steradian within the jet, $e_s$ is the energy per steradian within the remaining stellar material, and $\theta_c$ is the angle subtended by the material that is shock heated by the jet. These values are treated as free parameters. For our one-dimensional models of total explosion energy $E$, the energy per steradian $e$ is given by $e = E/4\pi$. Using two one-dimensional models with energy per steradian equal to a given $e_j$ and $e_s$, we average the chemical yields with a weighting determined by $\theta_c$.\\

This method for calculating the yields of aspherical supernovae is crude at best, but provides a cheap way to get some initial indication of the chemical yields we might expect from such aspherical supernovae, and allows us to examine how these yields might match the abundance profiles observed in metal-poor stars.

\section{Comparison With Observations}
We make use of \textsc{StarFit} \citep[\texttt{http://starfit.org},][]{chan_2016}, an evolutionary algorithm designed for abundance profiling, to search our dataset for the yields which best fit the observed abundances of a select number of metal-poor stars. The observational data that we use for comparison to our results are as follows. First, we compare our results with the average abundances of several metal-poor stars, all with [Mg/H]~<~-2.9 i.e. [Fe/H] $\lesssim -3$ \citep{cayrel_2004}. Also, we compare with HE~1300$+$0157, a CEMP star with [Fe/H] = -3.9. The abundance pattern of this star has been studied in detail by \citet{frebel_2007}. This star is found to be enriched in carbon and oxygen, but otherwise has abundances very similar to other EMP stars. We note that as part of the fitting algorithm, we allow for our yields to be shifted by a constant offset, i.e., the ratios between elements are held constant but the entire profile can be shifted up or down by a dilution factor to optimise the fit, as described by \citet{heger_2010}.

\subsection{Cayrel Data Set}
\begin{figure}
	\includegraphics[width=\columnwidth]{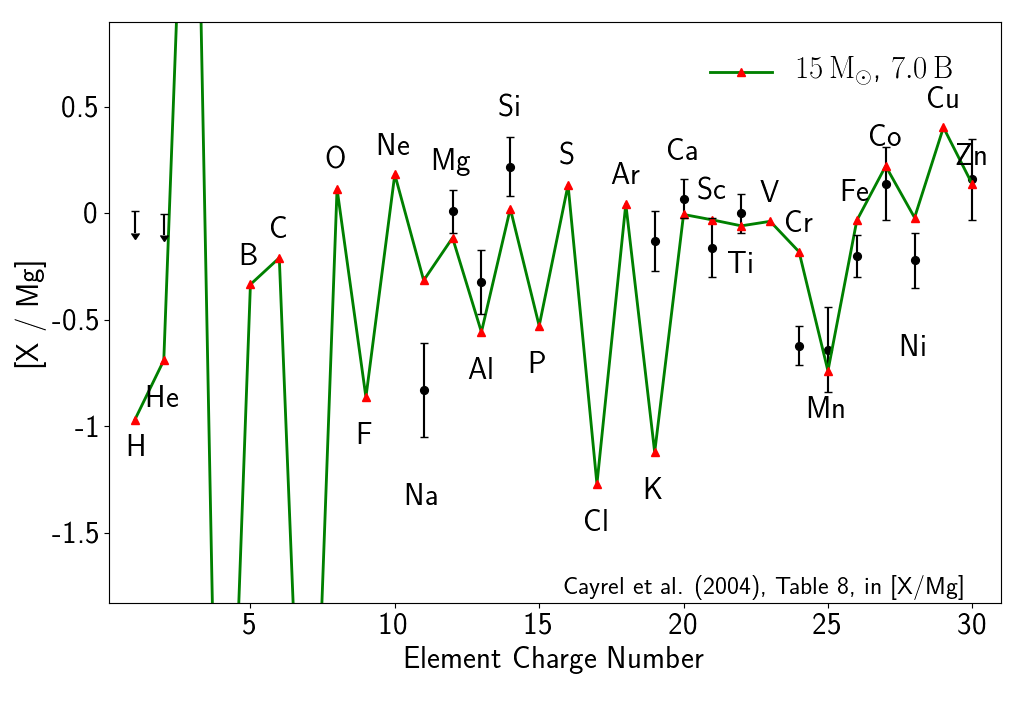}
    \caption{The abundance profile of the ejecta from a 7~B explosion of the $15\,\Msun$, plotted against the abundance profile of average low metallicity stars from the \citet{cayrel_2004} data set. Figure generated using \textsc{StarFit} \citep{chan_2016}.}
    \label{fig:starfit_cayrel2004}
\end{figure}

We find that the best matches for models of progenitor mass $\leq 30\,\Msun$ are models with explosion energies clustered around the value for which the reverse shock limits incomplete silicon burning at the inner shell interface. This is 6.5~B -- 7.5~B, 5~B -- 7~B, and 10~B -- 12~B, for the $15\,\Msun$, $20\,\Msun$ and $30\,\Msun$ models, respectively.\\
We find that a 7~B explosion of a $15\,\Msun$ progenitor, shown in Figure \ref{fig:starfit_cayrel2004}, provides the best fit to the averaged stars of lowest metallicity in the \citet{cayrel_2004} dataset. This model fits well to the observed abundance profile, and provides values for Ca, Sc, Ti, Mn, Co, and Zn all within observational uncertainties. Several elements are not so well fit. Of the light and intermediate mass elements, Na is overproduced, while all of our Mg, Al, Si, and K are underproduced. Though, of these, only K is more than $2\sigma$ from the observed value. Of the iron-peak, our Fe and Ni values are overproduced, but within $2\sigma$ of the observed values, while Cr is more than $3\sigma$ greater than the observed value. This particular model is one in which the reverse shock at the inner shell interface has a maximum effect on incomplete silicon burning, meaning that no other spherically symmetric model provides a better fit for Cr while still retaining reasonable values for other elements. The fit for K is improved for lower explosion energy, though the fit for the iron-peak is much worse for typical ($\sim 1$~B) explosions, as cobalt and zinc are significicantly underproduced. For models of progenitor mass $\geq 40\,\Msun$, we find that the models of best fit are those with very high explosion energy; greater than 20~B. In these models, low K and high Cr in comparison to observation remains a problem, and in addition, Fe begins to be overproduced.\\

\subsection{HE~1300$+$0157}
\begin{figure}
	\includegraphics[width=1.1\columnwidth]{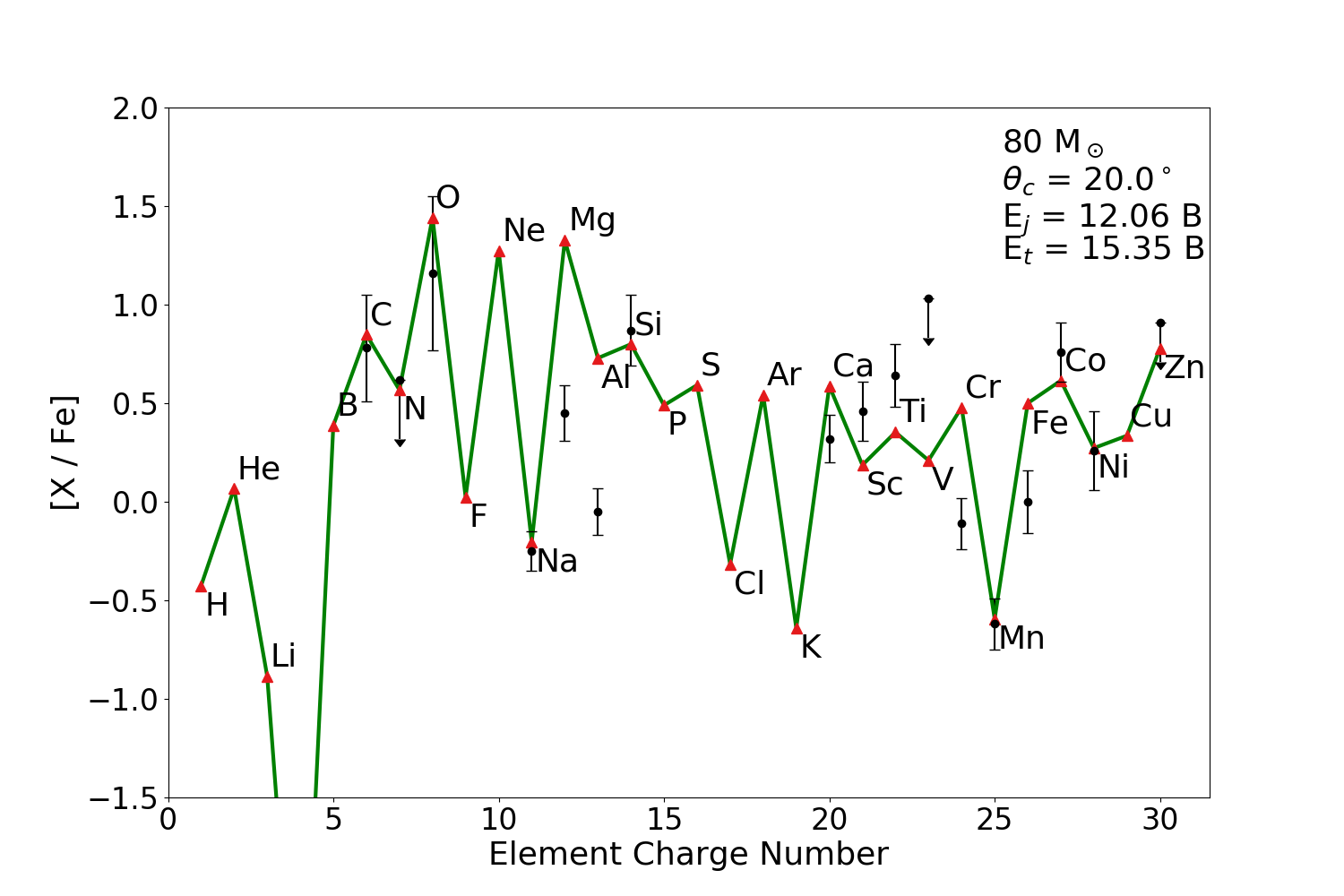}
    \caption{The estimated abundance profile of the ejecta from an aspherical explosion of the $80\,\Msun$ model compared to the observed abundances in the star HE~1300$+$0157 \citep{frebel_2007}. The total explosion energy is 15.35~B, where the jets carry 12.06~B of this energy. The half angle of the material shocked by the jet is 20$^\circ$.}
    \label{fig:starfit_HE13000157_z80}
\end{figure}

We are unable to reproduce the large carbon and oxygen abundances while still providing a reasonable match for the iron-peak abundances observed in this star with any of our spherically symmetric explosions. \citet{frebel_2007} and \citet{norris_2007} have proposed a scenario whereby these carbon enhanced metal poor stars may have been enriched by two sources, a faint supernova and a hypernova. The hypernova provides a good fit to iron peak elements, and a faint supernova with significant fallback of the inner layers may provide the carbon enhancement.
Alternatively, a similar enrichment may be achieved with an aspherical explosion, where much of the explosion energy is concentrated at the poles. The equatorial material carries just a small portion of the total explosion energy, and undergoes significant fallback \citep[e.g.,][]{tominaga_2007,tominaga_2009}.
We show that we can approximately reproduce the "carbon enhanced" abundance profile that is observed in this star with an estimate of such an explosion, as described in Section \ref{sec:jets}. As we see in Figure \ref{fig:starfit_HE13000157_z80}, we find a qualitatively good fit to the HE~1300$+$0157 using an $80\,\Msun$ model with total explosion energy slightly larger than 15~B, where approximately 12~B of this energy is carried by bipolar jets, which subtend a half angle of $20^\circ$. While the fit is far from perfect (e.g., the (Cr,Fe)/Mn ratio is far too large), this indicates that in principle, asymmetric supernovae may be a valid candidate for the enrichment sources of most iron poor of the carbon enhanced metal poor (CEMP) stars. The fraction of CEMP stars have been observed to increase with decreasing [Fe/H], and may be $>50\%$ for [Fe/H] $\lesssim-4$ \citep{cohen_2008,carollo_2012,lee_2013,norris_2013,yong_2013,hansen_2014}. If supernovae with bipolar jets are responsible for the enrichment of CEMP stars, then this would suggest that the ejecta from these explosions must have been dominant in the chemical enrichment of the early Universe. More work will need to be done in providing firmer constraints on the rotation speeds of the first stars, the explosion mechanism, and the nucleosynthesis that we should expect for bipolar explosions, to determine whether or not this could realistically be the case \citep[e.g.,][]{stacy_2011,stacy_2013a,macfadyen_1999,akiyama_2003,ardeljan_2005,burrows_2007,obergaulinger_2017}. Other suggestions for the origin of carbon-enhancement include enrichment via rotation-driven mass loss or mass transfer from a binary companion \citep{suda_2004,meynet_2006}, or self-enrichment driven by mixing \citep{campbell_2010}.

\section{Discussion}\label{sec:discussion}
We have calculated the nucleosynthetic yields for the spherically symmetric explosions of stars with initially zero metallicity. Our calculations are made for several different progenitor masses, between $15\,\Msun$ and $80\,\Msun$, and a wide range of explosion energies, covering at least 50 different energies between 0.05 B and 100 B for each progenitor model. We have also made a crude estimate for the yields of aspherical supernovae explosions, by integrating over a weighted grid of yields from one-dimensional models. \\

Our main objective has been to examine how the ejected abundances of the iron-peak elements vary with explosion energy, with the goal to gain insight into the nature of the first stars by comparing our results with the observed abundances of these elements in metal poor stars. In the following we discuss the implications of our results when compared to observed chemical abundances in iron poor stars.\\

We speculate that if the most iron poor stars in the \citet{cayrel_2004} dataset were to be enriched by a single supernova each, then the supernovae of primordial stars of progenitor mass $\leq30\,\Msun$ with explosion energy clustered in the range of approximately 5~B -- 10~B would be the best candidates to give a reasonable fit to the abundance ratios that are typically observed in these stars. \\
However, even in our models that fit very well to [(Mn,Co,Zn)/Fe] and most intermediate mass elements, we are greatly overproducing chromium. This is similar to the results found by both \citet{chieffi_2002} and \citet{heger_2010} in similiar studies, where [Cr/Fe] was systematically overproduced in comparison to the observed ratio in iron poor stars. This may be an indication that there needs to be some corrections made to our models. An increased neutron excess profile \cite[e.g.,][]{nakamura_1999} may increase the yields of manganese, cobalt and zinc, which all require a neutron excess, while leaving the iron and chromium yields unchanged, thereby deceasing the relative value of [Cr/Fe] in comparison to the other iron peak elements. It is also possible that the observational values of [Cr/Fe] suffer from some metallicity dependent corrections, as is suggested by \citet{cayrel_2004}, rather than an intrinsic decrease in [Cr/Fe] with metallicity, meaning that the true value of [Cr/Fe] in metal poor stars is greater than is commonly reported. This suspicion is raised by the very small scatter that is observed in values of [Cr/Fe] between stars of similar [Fe/H], which is in contrast to the scatter that one might expect to see in an abundance ratio that changes with metallicity, for example, as is observed in [Mn/Fe]. Futhermore, \citet{lai_2008} find an unexpected trend in the chromium abundances observed in some metal-poor stars with the effective temperature adopted for their models, which may suggest a discrepancy between the model atmospheres used to determine chemical abundances, and the true atmosheres in the stars being observed. \\
On the other hand, however, our results also show that [Cr/Fe] varies the least of all elements between models of different explosion energy, and also between models of different progenitor mass. There are only limited cases in which the [Cr/Fe] ratio fluctuates, i.e., when $\alpha$-particle production dominates the inner regions, and when silicon burning shifts out towards the inner shell interface, as discussed in \ref{ssec:alpharich} and Sections \ref{ssec:rev_shock}, respectively. Otherwise, chromium and iron production remain very tightly correlated across models, which may support the case that the variation in [Cr/Fe] is naturally very small between stars. In any case, even our lowest and best fitting values of [Cr/Fe] are too large to be compatible with the current observational data. \\

While we are able to achieve [(Co,Zn)/Fe] values compatible with those observed in metal poor stars using models of progenitor mass $\leq 30\,\Msun$ (and to some extent, the $40\,\Msun$ model) and an explosion energy of order 10 B, it is much more difficult to achieve high [(Co,Zn)/Fe] with the more massive progenitors. In both the $60\,\Msun$ and $80\,\Msun$ models, we can only achieve [(Co,Zn)/Fe] $\gtrsim 0$ with explosion energy of order 100 B. Although, we do find that we are able to achieve a reasonable fit to the carbon enhanced metal poor star HE~1300$+$0157 by approximating the yield of an aspherical explosion of the $80\,\Msun$ model. 
This may indicate that spherical explosions of stars more massive than $30\,\Msun$ -- $40\,\Msun$ either (i) mainly collapse into black holes and do not contribute to early chemical enrichment, (ii) mainly collapse into black holes, but a small amount of material that they do eject is mixed with the chemical enrichment from other nearby supernovae before forming the following generation of stars (as suggested by \citet{frebel_2007}), or (iii) explode by means other than core collapse. The outcome likely depends on whether rotation plays a role in the final fate of the star, as will be discussed later in this section.\\

Several studies on the chemical evolution of galaxies have included the yields from spherical hypernova models such as ours in their calculations, and found that the results were improved relative to GCE models without hypernovae to better match observed chemical abundances \citep{kobayashi_2006,nomoto_2006,nomoto_2013}. Typically, a free parameter is used to set the fraction of hypernovae to be included in the calculation. The explosion energy for the hypernova models in these studies is typically 10~B, with the regular supernova models having explosion energies of 1~B.\\
We have found that, particularly for model with progenitor mass  $\leq 30\,\Msun$, the ejected chemical abundance ratios can be very sensitive to the explosion energy and progenitor model structure, and the changes can be certainly non-linear between 1~B and 10~B. In our $15\,\Msun$ model, for example, an explosion energy of 7~B gives much lower [(Cr,Mn)/Fe] values than any other explosion energy. Likewise, the $20\,\Msun$ model gives much lowest [(Cr,Mn)/Fe] values at $\sim$6~B, and the $30\,\Msun$ model gives the lowest [(Cr,Mn)/Fe] at $\sim$10~B. This is not to suggest that the hypernova value should now be tuned to 5~B, or 7~B instead of 10~B in order to achieve the desired values of low [(Cr,Mn)/Fe], but it would be interesting to see what effect there is on GCE calculations by including some more continuous distribution of explosion energies for the first stars. We intend to explore idea this in a follow up study.\\

Similar to \citet{nakamura_1999,umeda_2002,umeda_2005,nomoto_2006} we have shown that, in theory, we are able to achieve a better fit to iron poor star abundances profiles using supernova models of much higher than typical explosion energy. The question remains, however, as to exactly how any exploding star would achieve such large explosion energy. It has been predicted that rotation and magnetic fields may provide an extra source of energy to during collapse \citep{leblanc_1970,bodenheimer_1983,macfadyen_1999}, however, these studies also predict that the effects of rotation will also be conducive to highly asymmetrical explosions. Retaining spherical symmetry is less computationally expensive, and has allowed us in this study to calculate the explosions and nucleosynthesis several hundred different models in order to gain a firmer understanding of the relationship between explosive nucleosynthesis and explosion energy. It would be useful, however, and perhaps more realistic, to extend this study to investigate the effects of aspherical explosions including full multidimensional hydrodynamics. \\
It has already been found that aspherical explosions, in particular those with bipolar jets, are capable of reproducing several of the chemical ratios observed in metal poor stars, such as enhanced [C/Fe], low [Mn/Fe] and increased [(Co,Zn)/Fe] \citep{maeda_2003,tominaga_2009}. It is also noted, however, that the ejected abundance ratios, along with the total amount of ejected $^{56}$Ni, is strongly dependent on the specific properties of the jets \citep{nagataki_2006,fujimoto_2007,ono_2009}. 
Here we have made a preliminary estimate of the yield from an aspherical explosion, and shown that we are able to approximately reproduce the [(C,O)/Fe] and [(Mn,Zn)/Fe] ratios commonly observed in metal poor stars, but a larger scale study into exactly how the ejected chemical abundances vary with jet properties is necessary to understand how aspherical supernovae may play a role in the chemical evolution of galaxies.\\

The exact mechanism by which a collapsing star is able to eject the outer layers with the large kinetic energy observed in supernovae remains unclear. In complex (computationally expensive) calculations, a successful explosion can be achieved by the inclusion of neutrino transport and multi-dimensional mixing effects; though not yet with much consistency \citep{herant_1994,burrows_1995,janka_1996,fryer_2000,muller_2012}. Inspired by the success of these multi-dimensional studies, a method has also been developed to achieve more robust neutrino driven explosions in spherical symmetry \citep{perego_2015}.
In simpler calculations such as ours, the explosion must be initiated by some more artificial means. In this case, we use a piston-driven explosion, as discussed in Section \ref{sec:method}. A similar method is the "kinetic bomb", where an outward velocity is imparted to an inner mass shell to generate a shock wave \citep[e.g.,][]{limongi_2012}. Yet another common method to initiate an explosion is with a "thermal bomb", whereby thermal energy is injected into the innermost regions of the star. The major differences that arise in the results of these separate methods are attributed to the hydrodynamical effects of injecting kinetic energy versus internal (thermal) energy. For equal explosion energy, piston driven/kinetic bomb models tend to accelerate inner material to comparatively higher velocities and lower temperatures, whereas thermal bomb models drive the inner material to higher temperatures but lower velocities \citep{aufderheide_1991}. Although the peak temperatures for each method converge during the later phases of the shock, the differences in the initial shock conditions strongly influence the final remnant mass and ejected chemical yields, particularly those isotopes which are synthesised in the early stages of shock, e.g., $^{56}$Ni \citep{young_2007}. \\

For example, \citet{tominaga_2007} also calculate supernova yields for a range of metal free progenitors, using the thermal bomb method to initiate the explosion. An artificial mixing and fallback effect is also included in their method. For the same explosion energy and progenitor mass, they find similar initial mass cuts, within $\pm0.1\,\Msun$ to those that we find. Their implementation of a mixing and fallback effect, however, increases their final remnant masses by up to $\sim1.5\,\Msun$. For lower explosion energy models, the amount of ejected $^{56}$Ni that they find is similar to the amount that we find for similar models, within a few percent. Though when comparing each of our large explosion energy models, we find up to 3 times more $^{56}$Ni in the ejecta, likely in part due to the much larger mass cuts that they take in their models.\\
\citet{limongi_2012} make a similar study, using a kinetic bomb to initiate the explosion. On average, they achieve remnant masses $\sim0.2\,\Msun$ larger than those that we find, and we typically eject a factor of $\sim$2 more $^{56}$Ni, for the same progenitor mass and similar explosion energy. These differences in remnant mass and ejected $^{56}$Ni are likely a result of the different locations that we each launch the shock. Although we both inject kinetic energy as a means to initiate the shock, \citet{limongi_2012} impart outward velocity to a shell at mass coordinate $\sim 1\,\Msun$, whereas we place our piston at approximately the base of the oxygen shell, as described in Section \ref{sec:method}. It may be the case that by the time the shock approaches the base of the oxygen shell in the models of \citet{limongi_2012}, a portion of the initial kinetic energy has been converted to thermal energy. This would result in comparitively lower velocities, causing the larger remnant masses and lesser amounts of ejected $^{56}$Ni.\\

It is very difficult to separate exactly which differences are a result of the chosen explosion method, as the situation is obscured by the differences in progenitor models, and in the case of \citet{tominaga_2007}, the additional mixing and fallback effect. There are several studies to which we could make quantitative comparisons \citep[e.g.,][]{nakamura_1999,umeda_2002,umeda_2005,heger_2010}, though to make a useful comparison with adequate consideration of the different progenitor models, explosion mechanisms, and various other initial conditions would require a separate study in itself. For the purposes of parameter space studies such as this one, it is probably sufficient that these simplified explosion models provide results which are in qualitative agreement with one another. Our results should serve to improve the current understanding of the relationship between supernova yields and explosion energy, while also providing some direction for future studies in this area. In particular, we highlight that care needs to be taken when including spherically symmetric hypernovae models in GCE calculations, as we show that the relationship between nucleosynthesis and explosion energy can be complex, and the exact choice of explosion energy for hypernova models should be given extra consideration. Furthermore, we emphasise the need for an improved understanding of the nucleosynthesis that we could expect from more realistic multidimensional hypernova models. While spherically symmetric hypernova models have enabled progress in our understanding of explosive nucleosynthesis and galactic chemical evolution, it is becoming clear that hypernovae are likely powered by rotation and magnetic fields, and therefore would deviate significantly from spherical symmetry.\\

All considered, the qualitative aspects of our results should be relevant no matter how exactly the shock is formed. For any strong shock travelling through a progenitor with an "onion skin" shell structure, the following effects occur;
\begin{description}
\item [(i)] Discontinuities in the $\rho r^3$ profile at shell interfaces will launch reverse shocks, which can increase peak temperatures by sufficient amounts to alter the burning process immediately interior to the shell interface. This effect is particularly strong when silicon burning is present at the outer edge of a shell, as the products are very sensitive to peak temperature.
\item [(ii)] Increasing the energy of the shock shifts and extends explosive nuclear burning regimes outwards, where changes in density and composition will alter the products of each burning regime, particularly those which require a neutron excess for synthesis (e.g., manganese, as $^{55}$Co).  
\item [(iii)] A stronger shock results in an increased degree of $\alpha$-rich freezeout. Depending on the density profile, or mass of the progenitor, increasing the explosion energy eventually causes $\alpha$-particle production to completely dominate the inner regions where iron-peak elements would otherwise be made.
\end{description}

\section{Conclusion}
In the context of the iron-peak elements, our results can be summarised as follows; an increase in explosion energy increases the mass fraction between complete silicon burning and incomplete silicon burning. This serves to increase [(Co,Zn)/Fe] and decrease [(Cr,Mn)/Fe], \textit{unless} the expansion time of inner material is such that $\alpha$-particle production dominates the final composition. In this case, the heavy element yields of complete silicon burning is reduced, undoing the effect of an increased mass fraction between the complete and incomplete silicon burning region, leading to an approximately constant [Cr/Fe] with explosion energy. [Mn/Fe] will typically continue to decrease, as manganese synthesis is limited by a decreased neutron excess at larger radius. For even greater explosion energy (faster expansion times), the extreme degree of $\alpha$-rich freezeout will reduce cobalt and zinc synthesis too, leading to a decrease in [(Co,Zn)/Fe] in this regime. For a narrow range of explosion energy in models $\leq 30\,\Msun$, reverse shock reduces the amount of incomplete silicon burning that occurs, and [(Cr,Mn)/Fe] are drastically reduced. \\

We find that the observed abundances of iron-peak elements [(Cr,Mn,Co,Zn)/Fe] are best reproduced in models where reverse shock plays the strongest role, typically for explosion energies between 5~B -- 10~B in models with progenitor mass $\leq 30\,\Msun$. Even these models, however, do not reproduce the entire typical abundance pattern observed in metal poor stars. Most notably, our [Cr/Fe] remains too high, and [K/Fe] too low to align with abundances oberved in the lowest metallicity stars. Spherically symmetric explosions of models with progenitor mass $\geq 40\,\Msun$ do not provide iron-peak yields that are compatible with metal-poor star observations, however, an approximation for the yields that these models would provide in aspherical explosions indicate that they may be a good candidate for the enrichment sources of those metal poor stars with enhanced carbon abundances. \\

\section*{Acknowledgements}
This work was supported by the Australian Research Council through ARC Future Fellowship FT160100035 (BM) and Future Fellowship FT120100363 (AH). AIK acknowledges financial support from the Australian Research Council (DP170100521). AH has been supported, in part, by a grant from Science and Technology Commission of Shanghai Municipality (Grants No.16DZ2260200) and National Natural Science Foundation of China (Grants No.11655002). This material is based upon work supported by the National Science Foundation under Grant No. PHY-1430152 (JINA Center for the Evolution of the Elements). This research was undertaken with the assistance of resources from the National Computational Infrastructure (NCI), which is supported by the Australian Government and was supported by resources provided by the Pawsey Supercomputing Centre with funding from the Australian Government and the Government of Western Australia. 




\bibliographystyle{mnras}
\bibliography{./james} 

\appendix

\section{Tables of Yields}

\input{z30_data.tex}


\bsp	
\label{lastpage}
\end{document}

%% file: z30_data.tex
\onecolumn
\begin{landscape}
\begin{longtable}{|c|c|c|c|c|c|c|c|c|c|c|c|c|c|c|c|c|c|c|c|c|c|c|c|c|c|c|c|c|}
\caption{An example of the yield tables available online. Here we show the yields in solar masses for the 30 solar mass model at 10 different explosion energies.}\\
\hline\hline\endhead
\hline\endfoot
Ion & 1 B & 2 B & 3 B & 4 B & 5 B & 6 B & 7 B & 8 B & 9 B & 10 B \\
\hline
H\ldots &    4.266e-01 &    4.266e-01 &    4.266e-01 &    4.266e-01 &    4.266e-01 &    4.266e-01 &    4.266e-01 &    4.266e-01 &    4.266e-01 &    4.266e-01\\
He\ldots &    3.261e-01 &    3.265e-01 &    3.270e-01 &    3.274e-01 &    3.278e-01 &    3.282e-01 &    3.286e-01 &    3.290e-01 &    3.294e-01 &    3.299e-01\\
Li\ldots &    2.834e-08 &    4.396e-08 &    6.082e-08 &    7.716e-08 &    9.250e-08 &    1.066e-07 &    1.197e-07 &    1.317e-07 &    1.426e-07 &    1.528e-07\\
Be\ldots &    2.002e-13 &    2.056e-13 &    2.026e-13 &    1.964e-13 &    1.885e-13 &    1.802e-13 &    1.723e-13 &    1.653e-13 &    1.592e-13 &    1.542e-13\\
B\ldots &    3.332e-08 &    2.654e-08 &    2.304e-08 &    2.094e-08 &    1.943e-08 &    1.825e-08 &    1.729e-08 &    1.649e-08 &    1.582e-08 &    1.525e-08\\
C\ldots &    1.037e-02 &    1.035e-02 &    1.033e-02 &    1.031e-02 &    1.029e-02 &    1.028e-02 &    1.026e-02 &    1.025e-02 &    1.024e-02 &    1.023e-02\\
N\ldots &    1.854e-04 &    1.844e-04 &    1.835e-04 &    1.828e-04 &    1.823e-04 &    1.818e-04 &    1.813e-04 &    1.809e-04 &    1.805e-04 &    1.802e-04\\
O\ldots &    1.749e-01 &    1.741e-01 &    1.738e-01 &    1.731e-01 &    1.722e-01 &    1.714e-01 &    1.706e-01 &    1.699e-01 &    1.692e-01 &    1.685e-01\\
F\ldots &    2.119e-06 &    1.768e-06 &    1.538e-06 &    1.372e-06 &    1.246e-06 &    1.145e-06 &    1.062e-06 &    9.927e-07 &    9.332e-07 &    8.817e-07\\
Ne\ldots &    3.907e-02 &    3.781e-02 &    3.674e-02 &    3.582e-02 &    3.500e-02 &    3.426e-02 &    3.359e-02 &    3.298e-02 &    3.242e-02 &    3.190e-02\\
Na\ldots &    1.320e-04 &    1.265e-04 &    1.219e-04 &    1.179e-04 &    1.144e-04 &    1.112e-04 &    1.083e-04 &    1.057e-04 &    1.033e-04 &    1.011e-04\\
Mg\ldots &    8.214e-03 &    8.321e-03 &    8.347e-03 &    8.366e-03 &    8.379e-03 &    8.387e-03 &    8.390e-03 &    8.387e-03 &    8.382e-03 &    8.376e-03\\
Al\ldots &    1.478e-04 &    1.504e-04 &    1.520e-04 &    1.535e-04 &    1.546e-04 &    1.556e-04 &    1.564e-04 &    1.570e-04 &    1.576e-04 &    1.581e-04\\
Si\ldots &    4.282e-03 &    4.515e-03 &    4.708e-03 &    4.988e-03 &    5.103e-03 &    5.283e-03 &    5.485e-03 &    5.776e-03 &    6.164e-03 &    6.578e-03\\
P\ldots &    9.799e-06 &    8.818e-06 &    8.557e-06 &    8.938e-06 &    9.248e-06 &    9.545e-06 &    9.829e-06 &    1.014e-05 &    1.053e-05 &    1.097e-05\\
S\ldots &    3.200e-03 &    3.207e-03 &    2.979e-03 &    3.099e-03 &    3.321e-03 &    3.436e-03 &    3.496e-03 &    3.600e-03 &    3.842e-03 &    4.215e-03\\
Cl\ldots &    4.255e-06 &    2.231e-06 &    1.960e-06 &    2.411e-06 &    2.638e-06 &    2.888e-06 &    3.080e-06 &    3.274e-06 &    3.475e-06 &    3.714e-06\\
Ar\ldots &    5.237e-04 &    5.506e-04 &    4.891e-04 &    5.208e-04 &    6.145e-04 &    6.497e-04 &    6.654e-04 &    6.701e-04 &    6.890e-04 &    7.507e-04\\
K\ldots &    3.217e-06 &    1.474e-06 &    8.434e-07 &    1.164e-06 &    1.298e-06 &    1.408e-06 &    1.483e-06 &    1.548e-06 &    1.609e-06 &    1.704e-06\\
Ca\ldots &    3.935e-04 &    4.563e-04 &    4.055e-04 &    4.446e-04 &    5.687e-04 &    6.121e-04 &    6.377e-04 &    6.383e-04 &    6.296e-04 &    6.673e-04\\
Sc\ldots &    1.775e-08 &    2.025e-08 &    3.730e-08 &    4.398e-08 &    8.855e-08 &    6.037e-08 &    3.895e-08 &    4.439e-08 &    6.686e-08 &    9.044e-08\\
Ti\ldots &    8.083e-06 &    1.002e-05 &    1.353e-05 &    1.535e-05 &    1.593e-05 &    1.732e-05 &    1.860e-05 &    1.916e-05 &    1.868e-05 &    1.890e-05\\
V\ldots &    7.908e-07 &    9.358e-07 &    1.046e-06 &    1.167e-06 &    1.182e-06 &    1.231e-06 &    1.312e-06 &    1.339e-06 &    1.306e-06 &    1.312e-06\\
Cr\ldots &    8.644e-05 &    9.753e-05 &    1.198e-04 &    1.338e-04 &    1.289e-04 &    1.416e-04 &    1.571e-04 &    1.583e-04 &    1.241e-04 &    1.011e-04\\
Mn\ldots &    2.441e-05 &    2.328e-05 &    2.401e-05 &    2.351e-05 &    2.195e-05 &    2.145e-05 &    2.077e-05 &    1.923e-05 &    1.523e-05 &    1.150e-05\\
Fe\ldots &    5.527e-03 &    6.678e-03 &    7.666e-03 &    8.372e-03 &    9.055e-03 &    9.768e-03 &    1.046e-02 &    1.098e-02 &    1.123e-02 &    1.105e-02\\
Co\ldots &    7.137e-06 &    1.131e-05 &    1.516e-05 &    1.796e-05 &    1.962e-05 &    2.103e-05 &    2.426e-05 &    2.787e-05 &    3.163e-05 &    3.460e-05\\
Ni\ldots &    1.670e-04 &    2.151e-04 &    2.564e-04 &    2.880e-04 &    3.193e-04 &    3.479e-04 &    3.726e-04 &    3.955e-04 &    4.201e-04 &    4.612e-04\\
Cu\ldots &    9.270e-07 &    2.387e-06 &    4.417e-06 &    6.043e-06 &    7.394e-06 &    8.503e-06 &    9.609e-06 &    1.071e-05 &    1.180e-05 &    1.324e-05\\
Zn\ldots &    4.416e-06 &    8.063e-06 &    1.287e-05 &    1.746e-05 &    2.044e-05 &    2.291e-05 &    2.497e-05 &    2.699e-05 &    2.898e-05 &    3.130e-05\\
Ga\ldots &    2.206e-09 &    2.768e-09 &    3.694e-09 &    4.497e-09 &    5.122e-09 &    5.739e-09 &    6.204e-09 &    6.617e-09 &    6.946e-09 &    7.261e-09\\
Ge\ldots &    1.291e-08 &    1.350e-08 &    1.561e-08 &    1.962e-08 &    2.233e-08 &    2.428e-08 &    2.593e-08 &    2.746e-08 &    2.868e-08 &    2.952e-08\\
As\ldots &    3.636e-12 &    5.470e-12 &    7.745e-12 &    1.087e-11 &    1.217e-11 &    1.325e-11 &    1.508e-11 &    1.734e-11 &    1.953e-11 &    2.124e-11\\
Se\ldots &    2.181e-10 &    2.347e-10 &    2.293e-10 &    3.037e-10 &    3.503e-10 &    3.829e-10 &    4.076e-10 &    4.303e-10 &    4.487e-10 &    4.594e-10\\
Br\ldots &    9.700e-14 &    1.850e-13 &    2.626e-13 &    3.666e-13 &    4.238e-13 &    4.488e-13 &    4.837e-13 &    5.386e-13 &    6.041e-13 &    6.679e-13\\
Kr\ldots &    4.261e-12 &    5.152e-12 &    4.485e-12 &    5.454e-12 &    6.507e-12 &    7.322e-12 &    7.981e-12 &    8.505e-12 &    8.931e-12 &    9.200e-12\\
Rb\ldots &    1.274e-17 &    1.790e-17 &    2.436e-17 &    2.920e-17 &    3.453e-17 &    3.879e-17 &    4.028e-17 &    4.110e-17 &    4.163e-17 &    4.333e-17\\
Sr\ldots &    1.013e-15 &    1.305e-15 &    1.537e-15 &    1.358e-15 &    1.586e-15 &    1.885e-15 &    2.065e-15 &    2.327e-15 &    2.592e-15 &    2.779e-15\\
Y\ldots &    2.666e-18 &    2.497e-18 &    2.319e-18 &    2.036e-18 &    2.014e-18 &    2.374e-18 &    2.719e-18 &    3.119e-18 &    3.432e-18 &    3.650e-18\\
Zr\ldots &    4.142e-18 &    6.239e-18 &    8.253e-18 &    8.352e-18 &    8.507e-18 &    1.079e-17 &    1.203e-17 &    1.339e-17 &    1.513e-17 &    1.684e-17\\
Nb\ldots &    9.396e-19 &    1.276e-18 &    1.574e-18 &    1.719e-18 &    1.575e-18 &    1.937e-18 &    2.305e-18 &    2.605e-18 &    2.984e-18 &    3.399e-18\\
\end{longtable}
\end{landscape}
\twocolumn